\documentclass[12pt,preprint]{aastex}

\usepackage{graphicx}
\usepackage{natbib}
\usepackage{color}
\usepackage{amsmath}

\newcommand*{\revision}{}

\newcommand*{\unit}[1]{\ensuremath{\mathrm{\, #1}}}

\newcommand*{\m}{\unit{m}}
\newcommand*{\km}{\unit{km}}
\newcommand*{\Mm}{\unit{Mm}}

\newcommand*{\pcmc}{\unit{cm}^{-3}}	
\newcommand*{\mmm}{\unit{m}^{-3}}           

\newcommand*{\ps}{\unit{s}^{-1}}   
\newcommand*{\pss}{\unit{s}^{-2}}   

\newcommand*{\kmps}{\km \ps}

\newcommand*{\mpss}{\m \pss}

\newcommand*{\g}{\unit{g}}
\newcommand*{\kg}{\unit{kg}}
\newcommand*{\gpcc}{\g \pcmc}    
\newcommand*{\kgpcm}{\kg \mmm}  

\newcommand*{\degree}{\ensuremath{^{\circ}}} 

\def \Ha {{H$\alpha$\ }}

\def \CaIIH {{\ion{Ca}{2}~H\ }}

\begin{document}

\title{Quiescent Prominence Dynamics observed with the \emph{Hinode} Solar Optical Telescope. II. Prominence Bubble Boundary Layer Characteristics and the onset of a coupled Kelvin-Helmholtz Rayleigh-Taylor Instability}

\author{Thomas Berger}
\affil{National Oceanic and Atmospheric Administration/National Weather Service\\
325 Broadway, Boulder, CO 80305}
\email{thomas.berger@noaa.gov}

\author{Andrew Hillier}
\affil{CEMPS, University of Exeter\\
Exeter EX4 4QF U.K.} 

\author{Wei Liu}
\affil{Bay Area Environmental Research Institute and Lockheed Martin Solar and Astrophysics Laboratory\\
3251 Hanover St., Palo Alto, CA 94304}

\begin{abstract}
We analyze solar quiescent prominence bubble characteristics and instability dynamics using \emph{Hinode}/Solar Optical Telescope (SOT) data. We measure bubble expansion rate, prominence downflows, and the profile of the boundary layer brightness and thickness as a function of time. The largest bubble analyzed rises into the prominence with a speed of about $1.3\kmps$ until it is destabilized by a localized shear flow on the boundary.  Boundary layer thickness grows gradually as prominence downflows deposit plasma onto the bubble with characteristic speeds of  $20 - 35\kmps$.  Lateral downflows initiate from the thickened boundary \revision{layer} with characteristic speeds of $25 - 50\kmps$, ``draining'' the layer of plasma. Strong shear flow across one bubble boundary leads to \revision{an apparent} coupled Kelvin-Helmholtz Rayleigh-Taylor (KH-RT) instability. We measure shear flow speeds above the bubble of $10\kmps$ and infer interior bubble flow speeds on the order of $100\kmps$. Comparing the measured growth rate of the instability to analytic expressions, we infer a magnetic flux density across the bubble boundary of $\sim\-10^{-3}$~T (10~gauss) at an angle of $\sim\-70\degree$ to the prominence plane. The results are consistent with the hypothesis that prominence bubbles are caused by magnetic flux that emerges below a prominence, setting up the conditions for RT, or combined KH-RT, instability flows that transport flux, helicity, and hot plasma upward into the overlying coronal magnetic flux rope.   
\end{abstract}

\section{Introduction}
\label{sec:intro}

The previous paper in this series \citep[][hereafter Paper~I]{Berger:2010bj} analyzed the kinematics of turbulent upflow plumes observed in quiescent  prominences by the Solar Optical Telescope \citep[SOT;][]{Tsuneta:2008be,Suematsu:2008gq} instrument on the Japanese/US/UK \emph{Hinode} \citep{Kosugi:2007ij} satellite. As hypothesized in Paper~I and by \cite{Ryutova:2010fd}, the upflow plumes, first identified in \cite{Berger:2008kx}, are generated by the magnetic Rayleigh-Taylor instability (RTI) taking place at the boundary between the prominence and dark ``bubbles" that rise into it. \cite{Leonardis:2012er} and \cite{Hillier:2017go} establish that small-scale velocity fluctuations in quiescent prominences are consistent with turbulent \emph{in situ} flows on the scale of the observed upflow plumes, implying that the RTI drives small-scale mixing in these structures. \cite{Freed:2016gw} use Fourier local correlation tracking \citep[FLCT,][]{Fisher:2008uf} to analyze quiescent prominence flows, including plumes, and establish power spectral densities for kinetic energy and vorticity. 

Numerical simulations have largely supported the RTI hypothesis for quiescent prominence plume formation. Hillier first simulated upflow plume formation using a model based on the analytical Kippenhahn-Schl{\"u}ter model with an embedded low-density region \citep{Hillier:2011hna,Hillier:2012fd,Hillier:2012bo} and found that plumes closely resembling those described in Paper~I were created by a magnetic RTI triggered by \revision{imposed} sinusoidal perturbations to the density inversion boundary. Hillier's simulations established that plume characteristics such as spatial and temporal scales and orientation are a function of gravity, the density inversion (quantified by the Atwood number $A = (\rho_{+} - \rho_{-})/ (\rho_{+} + \rho_{-})$ where $\rho_{+} > \rho{-}$ are the prominence and bubble densities, respectively), and the magnetic field strength and direction relative to the prominence/bubble boundary, consistent with analytical and numerical studies of magnetic RTI onset \citep[e.g.,][]{Chandrasekhar:1961a,Stone:2007jx,Stone:2007cw}. \cite{Keppens:2015iz} develop a 3D numerical simulation that includes the prominence-corona transition region \citep[PCTR;][]{Parenti:2012cc} and follows the resulting RTI into the fully non-linear regime to show that upflow plumes help maintain the prominence against gravitational drainage, as noted in the \revision{observations} of Paper~I.  \cite{Khomenko:2014br} further investigate partial ionization effects in a homogeneous magnetic field to  demonstrate a wider variety of plume and downflow dynamics, some of which replicate the ``droplet'' downflow formations analyzed in \cite{Haerendel:2011de}.  

In recent studies, \cite{Ruderman:2014cq} explore the effect of magnetic field shear on the RTI to show that it can modify the growth rate significantly but does not fully suppress the instability; \cite{Terradas:2015jn} investigate the effects of the plasma $\beta$ (the ratio of thermodynamic and magnetic pressures) and magnetic field line tying on the instability in a quiescent prominence/sheared arcade model; \revision{and \cite{Levens:2016kz} present vector magnetic field measurements of a prominence bubble boundary. \cite{Schmieder:2014kt} review open questions in contemporary prominence research}, \cite{Hillier:2017yy} provides a review of the RTI in solar prominences, \cite{Labrosse:2010bt} and \cite{Mackay:2010fp} provide recent reviews of prominence physics and modeling, \revision{and \cite{Vial:2015cn} present an extensive monograph on current prominence science}. 

Low (2007 private communication) first suggested \revision{to the authors} that prominence bubbles may be the manifestation of magnetic flux systems that emerge fortuitously beneath or just adjacent to prominences, e.g., as reported in \cite{Chae:2001jl}. In support of the emerging flux concept, \cite{Dudik:2012jy} and \cite{Gunar:2014he} show that the insertion of ``parasitic magnetic bipoles'' below linear force-free flux rope models of prominences \citep[e.g.,][]{Aulanier:1998tn} result in ``arcade'' structures that resemble the observed prominence bubbles.  Low further proposed that hot plasma, magnetic flux, and magnetic helicity can be transported upward by flows and instabilities from the emerging flux domain into the overlying coronal magnetic flux ropes, \revision{thus driving the flux ropes closer to instability and eventual eruption \citep{Low:2001vt}.} \cite{Xia:2016vs} model the quiescent prominence formation process numerically to confirm that hot plasma is transported upward from the chromosphere by dynamic flows, while radiatively cooled plasma returns to the chromosphere in the form of the ubiquitous vertical filamentary downflows \citep{Engvold:1981aa,Berger:2008kx}. Analysis of plasma motions in the ``coronal cavities'' associated with magnetic flux ropes \citep{Fuller:2009jm,Schmit:2011ix,Dove:2011gh,Li:2012hu,BakSteslicka:2013hg,Panesar:2013gu} and in the prominences themselves \citep{Okamoto:2009cg,Okamoto:2010gi,Okamoto:2016tw} support the proposition that flows from the lower atmosphere transport plasma and magnetic flux into system. If confirmed, small-scale emerging flux regions could represent a source term in the so-called ``chromosphere-corona mass cycle'' \citep{Marsch:2008aa,Berger:2011eoa,McIntosh:2012dx} in which hot plasma is transported upward into the corona while radiatively cooled plasma is channelled by magnetic fields to form visible cool plasma return flows in prominences \citep[e.g.,][]{Low:2012tu,Low:2012bj,Berger:2012da,Xia:2016vs} or coronal rain \citep{Antolin:2012jz}. In this context, prominence bubbles may be viewed as a mechanism that enables direct observation of upward transport flows in the coronal flux rope/prominence system. 

While there is general agreement that prominence bubbles are caused by emerging flux beneath pre-existing quiescent prominences, there is controversy surrounding the generating mechanism of the observed upflow plumes. \cite{Dudik:2012jy} and \cite{Gunar:2014he} assert that magnetic reconnection along the ``separator line'' tracing the bubble boundary drives the observed plume flows, while the work of Berger, Hillier, Keppens, Xia, and others demonstrate that the plumes are consistent with a magnetic RTI mechanism driven by the observed density inversion. In addition, while \cite{Berger:2011eoa} use differential emission measure (DEM) analysis to infer that the interior region of a prominence bubble contains plasma hotter than the surrounding chromosphere and corona in support of the ``magnetothermal convection'' hypothesis of Low, \cite{Dudik:2012jy} \revision{find} that in at least one case, a prominence bubble shows no enhancement in temperature over the background corona. \revision{Ground-based \Ha observations of prominence bubbles have sometimes shown bright core regions \citep{deToma:2008ei} or ``flare-like'' features \cite{Shen:2015gx} in bubbles which imply an internal heat source, but these observations are rare}. In any case, distinguishing between these ``hot bubble'' and ``empty arcade'' hypotheses on the basis of limb prominence observations alone is difficult due to the long line-of-sight through any foreground or background coronal plasma.  But differentiating between magnetic reconnection and RTI mechanisms in generating the upflow plumes can be addressed through examination of the bubble boundary layer region from which the plumes originate. 

To that end, in this paper we analyze two typical prominence bubbles and focus on bubble growth rates, bubble boundary layer characteristics, and the influence of shear flows on bubble stability. We show that shear flows on the bubble boundary can apparently lead to Kelvin-Helmholtz instability (KHI) flows that modify the upflow plume dynamics. We assume that the subsequent upflow plumes are driven by the RTI and use analytical studies of the coupled KH-RT instability to infer prominence and bubble magnetic field characteristics that are consistent with vector magnetic field measurements of other quiescent prominences. Coupling between the KH and RT instabilities has been studied theoretically \citep{Guzdar:1982gz,Zhang:2005it,Olson:2011ht,Ye:2011gl} but has not, to our knowledge, been observed in a real astrophysical system. We analyze the growth rate and boundary dynamics of a large prominence bubble boundary in Section~\ref{sec:growth}. In Sec.~\ref{sec:shear} we analyze the clearest case of sustained shear flow along a bubble boundary in the SOT database to determine the flow and instability parameters. Sec.~\ref{sec:khrt} derives implications for the prominence magnetic field under the assumption that we are observing a coupled KH-RT instability. Finally, we summarize our findings and conclusions in Sec.~\ref{sec:summary}. Details of instrument characteristics and data preparation for SOT prominence studies can be found in previous publications \citep[e.g., Paper~I and][]{Chae:2008ca}.

\section{Prominence Bubble Expansion and Boundary Layer Characteristics}
\label{sec:growth}

The following analysis focuses on the largest prominence bubble event in the SOT database: the 16-August-2007 bubble event that was observed continuously in the Ca~II H-line bandpass for over six hours with a mean cadence of 25~s (see online Movie~1). \revision{We} measure the bubble rise speed,  characterize the bubble boundary layer, and assess the associated flows into, and from, the boundary region as the bubble expands into the prominence above.  This particular prominence bubble was unusual in that it did not produce a series of smaller RT plumes as it evolved, but instead grew to the very large height of approximately 25~Mm above the photosphere. The bubble then appears to ``pop'' and release one large RT plume that travels completely through the prominence above, presumably into the overlying coronal magnetic flux rope and associated coronal cavity.   However the size of the bubble and the typical appearance of its boundary layer make this event ideal for studying the details of boundary flows in these structures. 

\begin{figure} 
\begin{center}
\includegraphics[width=0.85\textwidth]{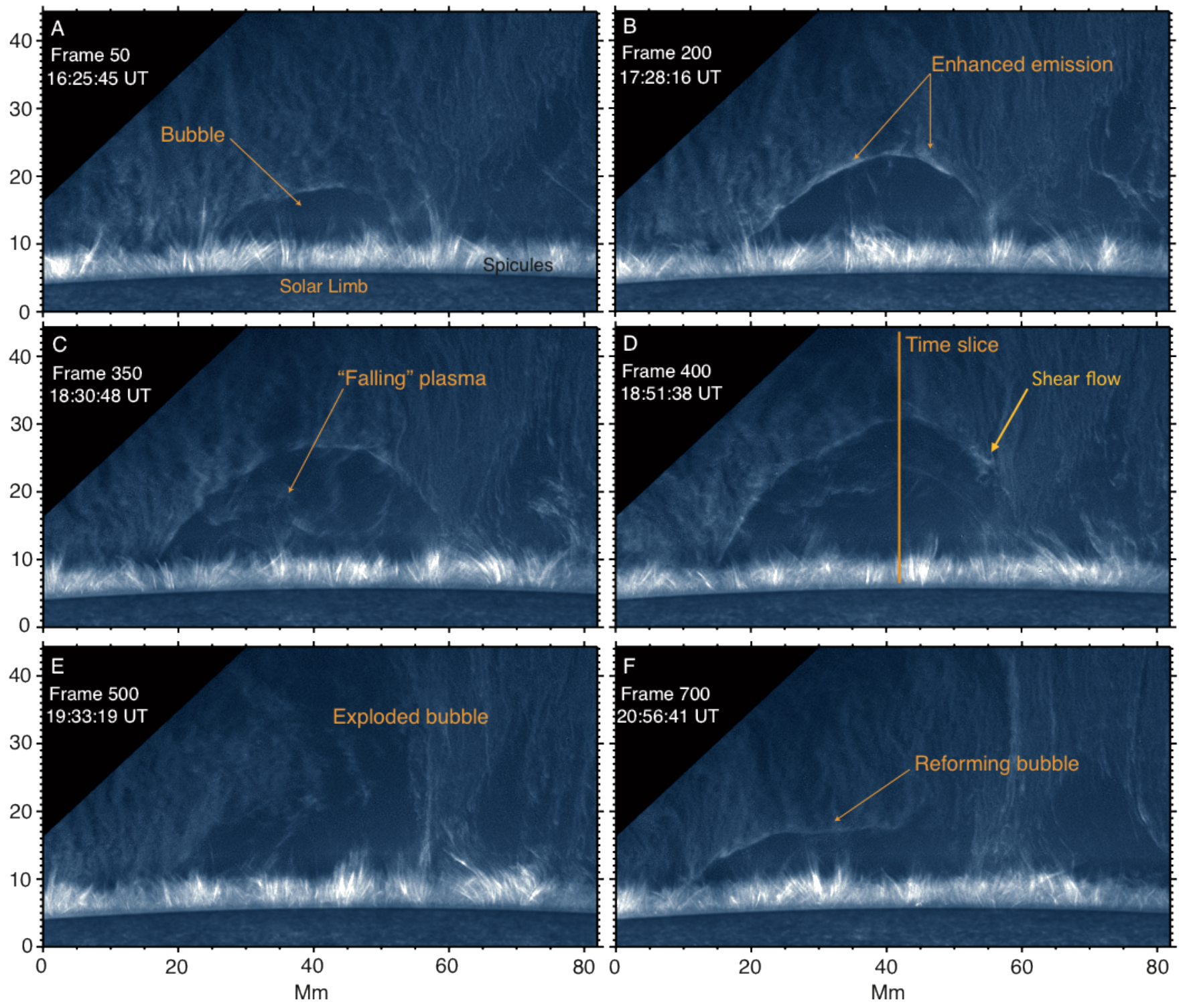}
\caption{Quiescent solar prominence observed on 16-August-2007 in the SOT Ca~II H-line 396.8~nm channel at 90W 47N heliographic coordinates. The missing data in the upper left corner of the images is outside the SOT field-of-view (FOV). The pixel scale is 0.079 Mm pixel$^{-1}$ and the image has been rotated to place the solar limb horizontal. Axes are labelled in Mm. The orange line marked "Time slice" in Panel~D shows the location of the space-time images shown in Figures.~\ref{fig:16aug2007_timeslice}, \ref{fig:16aug2007_timeslice_closeup}, and \ref{fig:16aug2007_timeslice_2}. A movie accompanying this figure is available in the online journal (movie1.mov).}
\end{center}
\label{fig:16aug2007_main}
\end{figure}

Figure~\ref{fig:16aug2007_main} shows a series of images of the 16-August-2007 quiescent prominence. At about 16:00~UT a bubble rises from the spicule foreground and by Frame~50 (16:25 UT, Panel~A) of the time series has formed into a roughly semi-circular void with a radius of about 10~Mm.  By Frame~200 (17:28 UT, Panel~B) the bubble has grown to a height above the photospheric limb of about 15~Mm and extended on the Northern (left) side to develop a long sloping boundary. Panel~B also shows areas of brighter emission along the bubble boundary where downflowing plasma from the prominence above has collected. Frame~350 (18:30 UT, Panel~C) shows plasma falling from the bubble boundary in large cascades that cause a decrease in the areas of bright emission pointed out in Panel~B. This ``falling plasma" is frequently observed in large prominence bubbles like this one, \revision{however we note that these flows are not correlated with the production of upflow plumes -- i.e., they do not indicate the onset of the RTI.} Frame~400 (18:51UT, Panel~D) shows the bubble in its late stage of evolution with steepened sides on both the left and right side boundaries. At the location indicated in Panel~D, the onset of a strong, nearly vertical, shear flow appears to rapidly destabilize the bubble boundary, leading to the formation of a single large upflow plume (characteristic of a high Atwood \revision{number} instability) that essentially ``explodes'' the bubble around 19~UT. By Frame~500 (19:33 UT, Panel~E) the bubble boundary has disintegrated and the region is in the process of being filled in by plasma from the prominence above. Finally, in Frame~700 (20:56 UT, Panel~F) the bubble is shown reforming in essentially the same location but with a more distorted and flattened profile. The SOT time series ends before the reformed bubble reaches instability onset. 

The 16-August-2007 prominence bubble illustrates many of the common traits seen in SOT quiescent prominence observations. Those traits are (a) the emergence of a semi-circular or elliptical dark void that pushes up into the prominence from below; (b) steady brightening on the bubble boundary at locations where downflows of prominence plasma intersect the boundary; (c) steepening of the bubble boundary and/or shear flows that cause perturbations in the boundary; and (d) plume formation initiated at the site of the boundary perturbations. In some cases formation of a large RT plume causes rapid collapse of the bubble (e.g. the 16-August-2007 case shown above) while in others the bubble can remain stable for hours while releasing multiple smaller plumes (e.g. the 30-November-2006 event analyzed in Paper~I). The Atwood number of the prominence-bubble system and the magnitude of the shear flows along the boundary are likely contributing factors to the scale of the eventual RT instability flows. 

\begin{figure} 
\begin{center}
\includegraphics[width=0.75\textwidth]{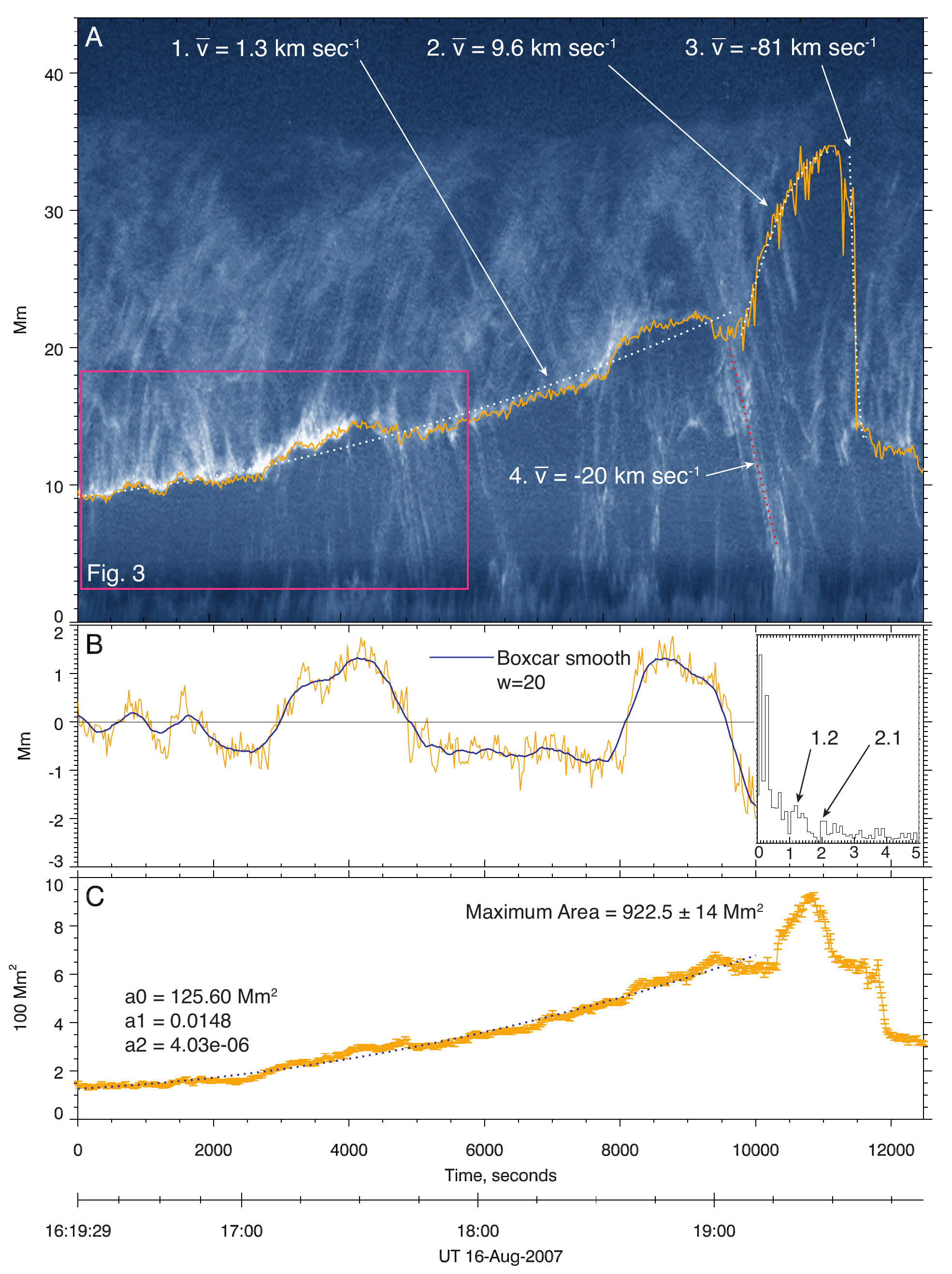}
\caption{(A) Space-time analysis of 16-August-2007 bubble boundary trajectory. The plot is generated along the vertical line in Fig.~\ref{fig:16aug2007_main}D and shows parabolic fits ($y = a_{0} + a_{1}t + a_{2}t^{2}$) to boundary trajectories (Curves 1--3) and a fit to the large downflow initiating around 19:00~UT. The orange boundary curve is defined by the intersection of the vertical line in Fig.~\ref{fig:16aug2007_main}D with the manually defined bubble shape. The red box shows the area of detail analyzed in Fig.~\ref{fig:16aug2007_timeslice_closeup}. (B) Plot of boundary profile with a parabolic fit (dotted Curve~1 in Panel~A) subtracted to show normalized departures from the mean trajectory. Inset shows a Fourier frequency analysis of the normalized trajectory in units of power (squared amplitude) vs. mHz. (C) Bubble area growth rate plot with 2nd order least-squares fit shown by the black dotted line. Values $a_{i}, i=0,1,2$, indicate the parabolic fit parameters.}
\end{center}
\label{fig:16aug2007_timeslice}
\end{figure}

Figure~\ref{fig:16aug2007_timeslice} shows a space-time analysis of the growth of the bubble along the line shown in Fig.~\ref{fig:16aug2007_main}D, approximately where the right half of the bubble explodes upward between Frames 400 and 500. Panel~A measures the velocities associated with both the bubble growth and collapse (curves 1--3) as well as one of the major downflows that occur just prior to the major instability onset at approximately 19:00~UT (curve 4). The bubble boundary trajectory is computed as the intersection of the vertical line in Fig.~\ref{fig:16aug2007_main}D with the bubble shape traced manually in each frame of the time series. Manual tracing of the bubble shape in each frame of the time series was done with a pen tablet device with an estimated precision in the boundary determination of $\pm5$~pixels or about $\pm0.4$~Mm. This accounts for the majority of the high-frequency noise in the normalized plot of Fig.~\ref{fig:16aug2007_timeslice}B. 

Curves 1 and 2 in Fig.~\ref{fig:16aug2007_timeslice} are parabolic fits to the bubble boundary trajectory showing that the mean speed during the slow rise phase is 1.3~$\kmps$, accelerating to 9.6~$\kmps$ during the non-linear RTI phase. Curve 3 with a mean speed of $-81~\kmps$ is a fit to the discontinuous jump of the bubble boundary after the large-scale instability onset and therefore is not necessarily indicative of a mass flow -- it more likely indicates where the boundary of the bubble can first be detected after the downflow plasma is no longer obscuring it. Just prior to instability onset, a large downflow of plasma demarcated by Curve~4 occurs at $-20~\kmps$, approximately two times faster than the typical downflow speed measured in \cite{Berger:2008kx} but comparable to speeds measured in \cite{Chae:2010ds} and \cite{Liu:2012gqa}, for example. 

Fig.~\ref{fig:16aug2007_timeslice}B shows the slow rise phase boundary trace with the parabolic fit shown in Panel~A subtracted. This plot accentuates the two episodes of sudden vertical ``jumps'' at the boundary occurring at about 17:00 and 18:30~UT. It also demonstrates the magnitude of the boundary measurement precision as evidenced by the high-frequency variation around the smoothed curve (produced using a ``boxcar'' window of 20 time-steps in width). The inset shows a Fourier power spectrum analysis of the boundary trajectory after subtraction of the smoothed curve, searching for characteristic oscillation frequencies in the time series. There are weak indications of possible peaks at 1.2 and 2.1~mHz, but in general the plot is consistent with random variation; we find no evidence for photospheric pressure-driven oscillations in the motion of this bubble boundary. Panel~C shows a plot of area growth rate showing a parabolic fit to the first 10000~s of evolution. The bubble reaches a maximum sky-plane-projected area of 922.5$\pm$14~$\Mm^2$ just prior to instability onset and collapse. 

\begin{figure} 
\begin{center}
\includegraphics[width=0.75\textwidth]{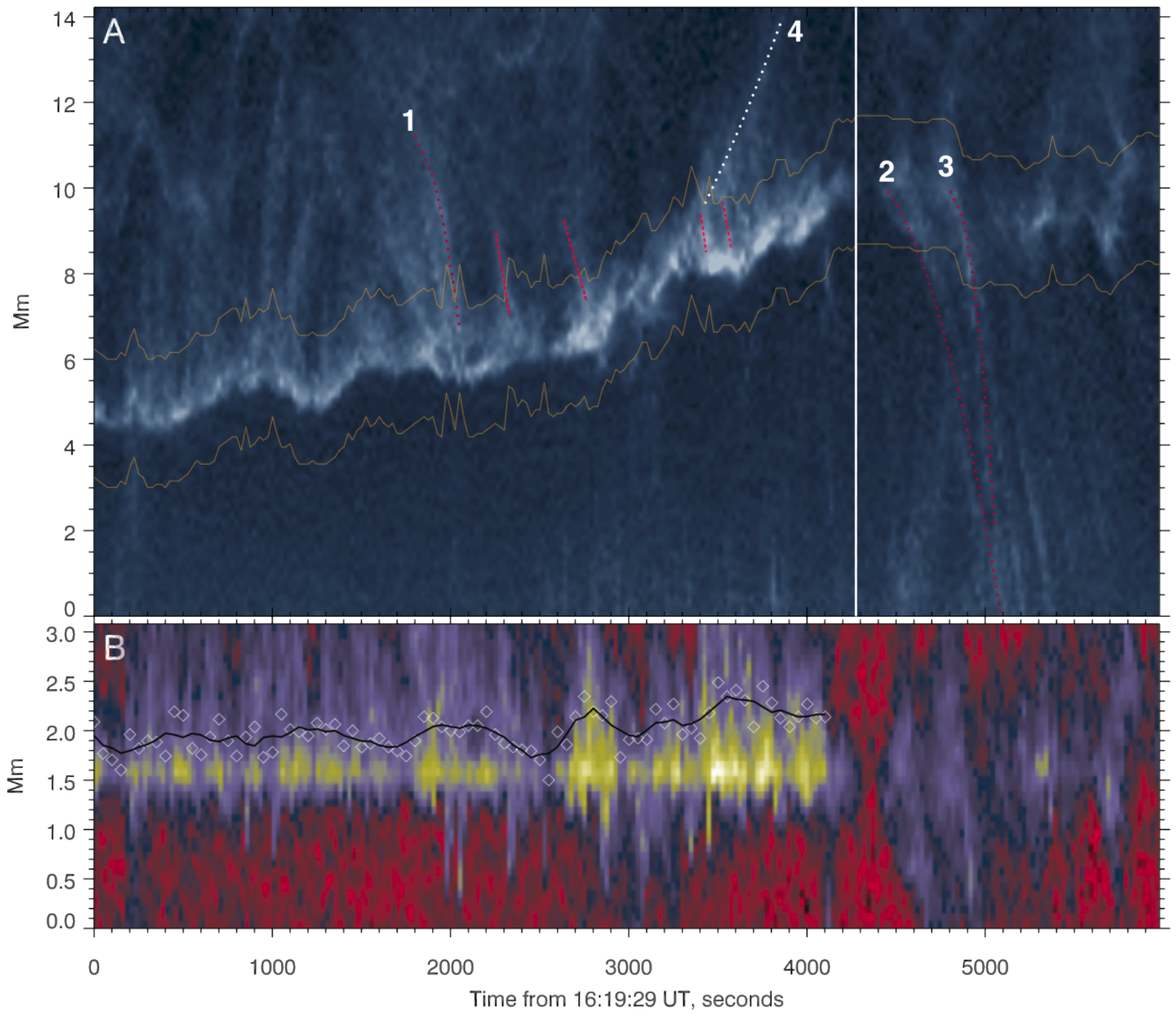}
\caption{(A) Detail of 16-August-2007 bubble boundary area indicated by the red box in Fig.~\ref{fig:16aug2007_timeslice}. \revision{Ordinate units are plane-of-sky height difference in Mm}. The yellow traces around the boundary indicate the area analyzed in Panel~B. Four flows are demarcated in the figure: Red dashed curves are downflows that occur above (Curve 1 and shorter curves) and below (Curves 2 \& 3) the boundary. Curve 4 is an apparent upflow of plasma from the boundary due at least partially to lateral transport across the time-slice (see text). The vertical white line demarcates a break in the boundary thickness buildup caused by the downflows indicated by Curves 2 \& 3.  (B) Intensity plot of the boundary area between the yellow lines in Panel~A created by stacking vertical slices that are then straightened along the central path. The color table scales on brightness of the boundary. The overplotted diamond symbols every 50~s indicate the time history of boundary brightness (in arbitrary units) up to the time marked by the white line in Panel~A. The black curve is a 5-point box-car smoothed display of the brightness data as a function of time.}
\end{center}
\label{fig:16aug2007_timeslice_closeup}
\end{figure}

Figure~\ref{fig:16aug2007_timeslice_closeup} shows an enlarged version of the bubble boundary space-time image. Panel~A shows that plasma draining down from the prominence, for example the flow marked by Curve~1, is the primary source of the increased boundary brightness. The 2nd-order least-squares fit to the downflow marked by Curve~1 indicates an acceleration of approximately $60~\mathrm{m\ s}^{-2}$ (22\% of solar gravitational acceleration) to a final speed of about 35~$\kmps$ just before it terminates in the bubble boundary. Such downflows appear to be the main source of bubble boundary buildup over time. But the buildup does not continue indefinitely. In this case, approximately 4000~s from the start of the time series (near 17:35~UT) the boundary suddenly decreases in brightness, apparently due to the generation of at least two large downflow streams highlighted as Curves~2 and 3 in Fig.~\ref{fig:16aug2007_timeslice_closeup}. Second order polynomial fits to Curves~2 and 3 show that these ``drainage downflows'' accelerate at approximately $10~\mathrm{m\ s}^{-2}$ to achieve sky-plane speeds of $26~\kmps$ and $48~\kmps$, respectively.  The low acceleration values are likely due to significant out-of-sky-plane velocity components that are not captured in these curve fits. 

Finally, Curve~4 in Fig.~\ref{fig:16aug2007_timeslice_closeup} traces an upflow that appears to commence from the boundary at about 17:10~UT. The 2nd-order fit to this trajectory indicates a final speed of approximately $11.6~\kmps$. Although the speed of this upflow is similar to that of RT plumes measured in other prominences \citep[e.g.,][]{Berger:2010bj}, the origin of these upflows appears to be plasma that is being pushed upwards along the bubble to intersect the time-slice plane. Similarly, the very steep apparent upflow curves seen around 1000~s in the time slice likely do not represent vertical plasma motions in the time-slice plane but are lateral flows along the bubble boundary. Similar lateral upflows in quiescent prominences are analyzed in \cite{Chae:2008ca}.

Figure~\ref{fig:16aug2007_timeslice_closeup}B shows a stack plot of vertical slices through the boundary between the yellow curves in Panel~A. The color map emphasizes the boundary brightness as a function of distance (y-axis) from the center of the boundary line at a given time (x-axis). The boundary brightness is relatively stable until about 2500~s in the time series, at which time it grows with some oscillatory behavior to a maximum level just prior to the drainage event. This growth is directly correlated with the deposition of plasma into the boundary layer by the downflows marked in Panel~A. The plot establishes that boundary brightness is correlated with the thickness of the layer with the dimmer boundary values corresponding to thicknesses of approximately 0.25~Mm and the brightest values corresponding to thicknesses across the boundary of approximately 0.5~Mm. This is the behavior expected of an optically thin plasma in emission against a dark background. There is no evidence of impulsive brightening of the boundary that is not correlated with prominence plasma downflows. We see no evidence of sudden brightening and flow triggering that would be expected in cases of magnetic reconnection at the boundary. 

We emphasize that the acceleration and velocity values presented here are measured in a vertical time-slice which is itself taken in a sky-plane projected image series -- the magnitudes quoted do not necessarily apply to vertical motions aligned along the solar gravity vector. In fact, the large downflows that appear to drain the boundary at around 17:35~UT have distinctly ``arched'' trajectories (see Movie~1) that may indicate that the flows are sliding around the bubble rather than through it. Multiple viewpoints of a prominence bubble event would enable a more thorough analysis of plasma flows on the boundary, enabling a determination of the true directions of the observed flows relative to the prominence and bubble structures. 

\begin{figure} 
\begin{center}
\includegraphics[width=0.5\textwidth]{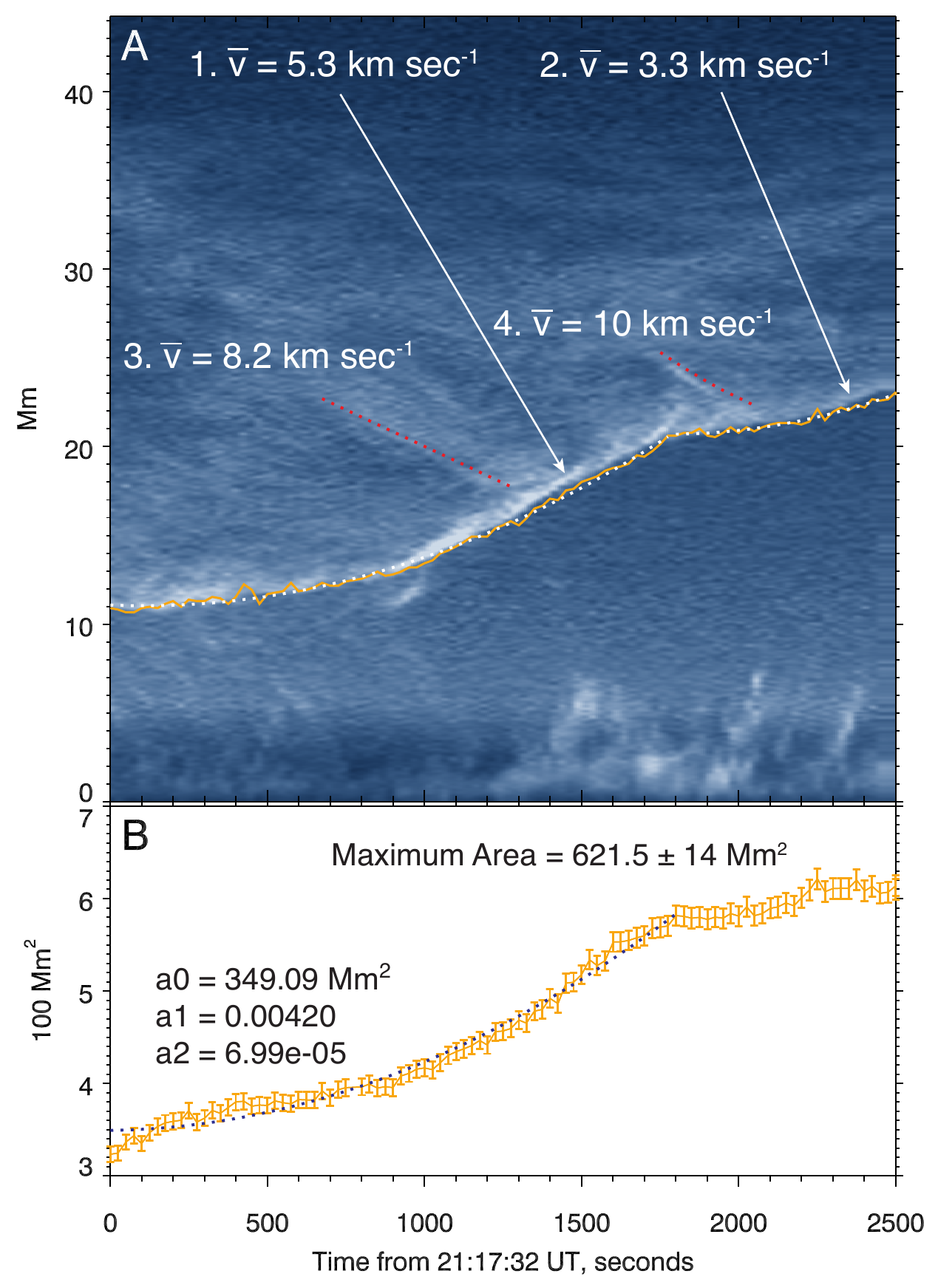}
\caption{(A) Space-time analysis of the 2nd bubble boundary trajectory in the 16-August-2007 prominence. The plot is generated along the vertical line in Fig.~\ref{fig:16aug2007_main} Panel~D. (B) Recurring bubble area growth rate plot with 2nd order polynomial fit shown by the black dotted line. Values $a_{i}, i=0,1,2$, indicate the fit parameters.}
\end{center}
\label{fig:16aug2007_timeslice_2}
\end{figure}

Following the collapse of the bubble around 19:30~UT there is a period where the boundary cannot be clearly defined in the images. It appears from the movie that the region under the prominence remains ``magnetically influenced'' even though the boundary region is ragged and poorly defined. However by about 21:00~UT another well-defined bubble builds back into the prominence from the same location as the previous bubble. The growth rate of the ``recurring bubble'' is shown in Fig.~\ref{fig:16aug2007_timeslice_2} as a space-time slice identical in construction to that in Fig.~\ref{fig:16aug2007_timeslice}. In the second bubble formation event the growth rate is significantly faster, with a boundary mean vertical rise speed of 5.3~$\kmps$ for at least 1800~s of growth. The rise speed then decreases to a mean of 3.3~$\kmps$ for the remainder of the observation period. The recurring bubble has approximately the same lateral extent as the first bubble shown in Fig.~\ref{fig:16aug2007_main}, demonstrating that growth rates are not necessarily related to bubble size. Small downflows occur frequently at the beginning of the growth period of the second bubble and then stop after about 1000~s of evolution.  Distinct prominence downflows (Curves~3 and 4 in the figure) with speeds of 8.2 and 10~$\kmps$ contribute to the boundary growth during this period. Fig.~\ref{fig:16aug2007_timeslice_2}B verifies the accelerated growth rate of the bubble area compared to the first period, but the projected area plateaus at a value of about 620~Mm$^2$. It is likely that this bubble continued to grow until it became unstable, but the observation period ends just after the plateau in growth rate.

Finally, we find no observable unusual activity in the chromosphere below the bubble region that correlates with the bubble rise or instability phases. For example, the spicules in line with the bubble outline in Movie~1 show no unusual enhancements in brightness or magnitude during the first or second rise phases or just prior to the instability onset around 19~UT.

\section{Boundary Layer Shear Flow Characteristics}
\label{sec:shear}

In the bubble event analyzed above, the collapse is apparently triggered by an instability due to increased shear flows on the right boundary of the bubble around 18:50~UT, as previously noted by \cite{Ryutova:2010fd}. The onset of the instability in this case is rapid and localized, and as a result is difficult to analyze. In this section, we examine shear flows on a smaller bubble that went unstable over a longer period and was observed with 10~s temporal cadence, allowing more detailed analysis of the flow and instability characteristics. This particular bubble has also been used in past studies to illustrate prominence plume dynamics \citep[e.g.,][]{Berger:2010bj, Hillier:2012fd}. 

\begin{figure}  
\begin{center}
\includegraphics[width=0.8\textwidth]{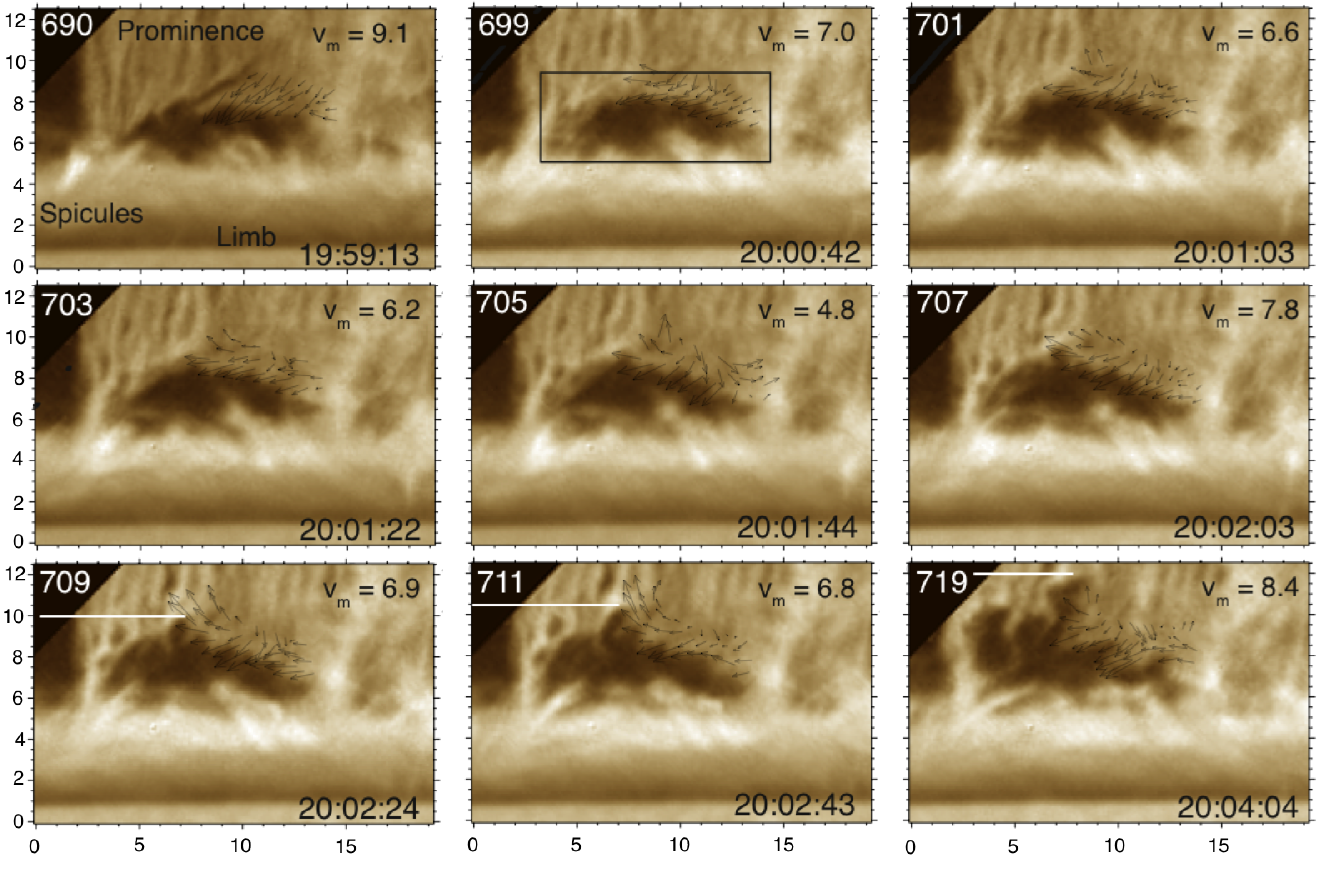}
\caption{Bubble evolution during 08-August-2007 event with prominence plasma flow vectors overplotted in the region where significant shear flow develops. Frames are numbered in the upper left from the start of the observation period. The pixel scale is 0.116 Mm pixel$^{-1}$ and the image has been rotated to place the solar limb horizontal. Axes are labelled in Mm \revision{height from an arbitrary reference level}. $\mathrm{v}_\mathrm{m}$ ($\kmps$) in each panel is the mean speed of the flow vectors at each time step, labelled with UT time in the bottom right. The black box in Frame~699 is shown enlarged in Fig.~\ref{fig:0808_bumps}. The white line in the last three panels demarcates the top of the first plume to generate from the boundary. The black triangular areas in the upper left are regions outside the SOT FOV for this dataset. Tickmarks are Mm. A movie accompanying this figure is available in the online journal (movie2.mov).}
\end{center}
\label{fig:0808_montage}
\end{figure}

Figure~\ref{fig:0808_montage} shows a series of \Ha line-center filtergrams of the small prominence on the NE limb of the Sun observed by the SOT on 08-August-2007. The full time series spans four hours from about 18 to 22~UT with a mean cadence of 10~s and a detector spatial scale of 0.116~Mm/pixel (0.16~arcsec/pixel). Here we concentrate on the short period of time surrounding the initial plume formation, approximately 4~minutes prior to instability onset. The generating prominence bubble is relatively small, achieving a maximum height before instability onset of only about 8~Mm above the photosphere and about 3~Mm above the tops of the chromospheric ``spicules'' that reach 5--6~Mm above the photospheric limb. Nevertheless, this event was observed with high temporal cadence and offers one of the best data series for studying the flow dynamics surrounding RTI onset in solar prominences. 

The plasma flows near the upper boundary of the prominence bubble are measured using the NAVE optical flow tracking code \citep{Chae:2008jq} on the full 10~s cadence data set with a grid spacing of 0.5~Mm. The resultant flow vectors are plotted in Fig.~\ref{fig:0808_montage} as black arrows in a select region that follows the upper right boundary of the bubble. The highest velocity arrow in each frame is normalized to 1.75~Mm in length on the image using the IDL velovect.pro routine with a length parameter of 15 pixels. The \revision{mean velocity, $V_m$,} of all plotted velocity vectors is shown in each frame of the figure. 

\begin{figure}   
\begin{center}
\includegraphics[width=0.485\textwidth,trim=2mm 0mm 1mm 0mm, clip]{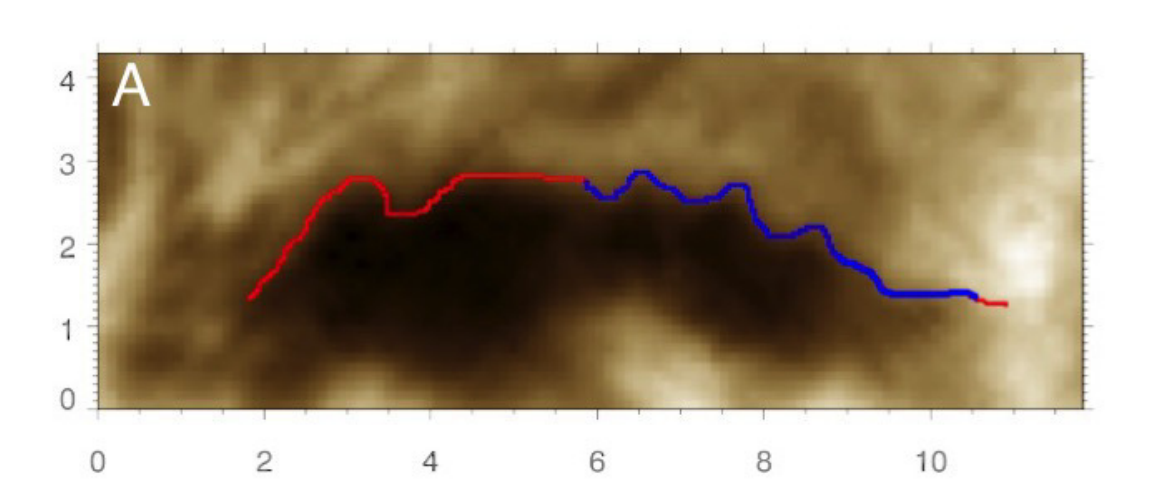}
\includegraphics[width=0.495\textwidth]{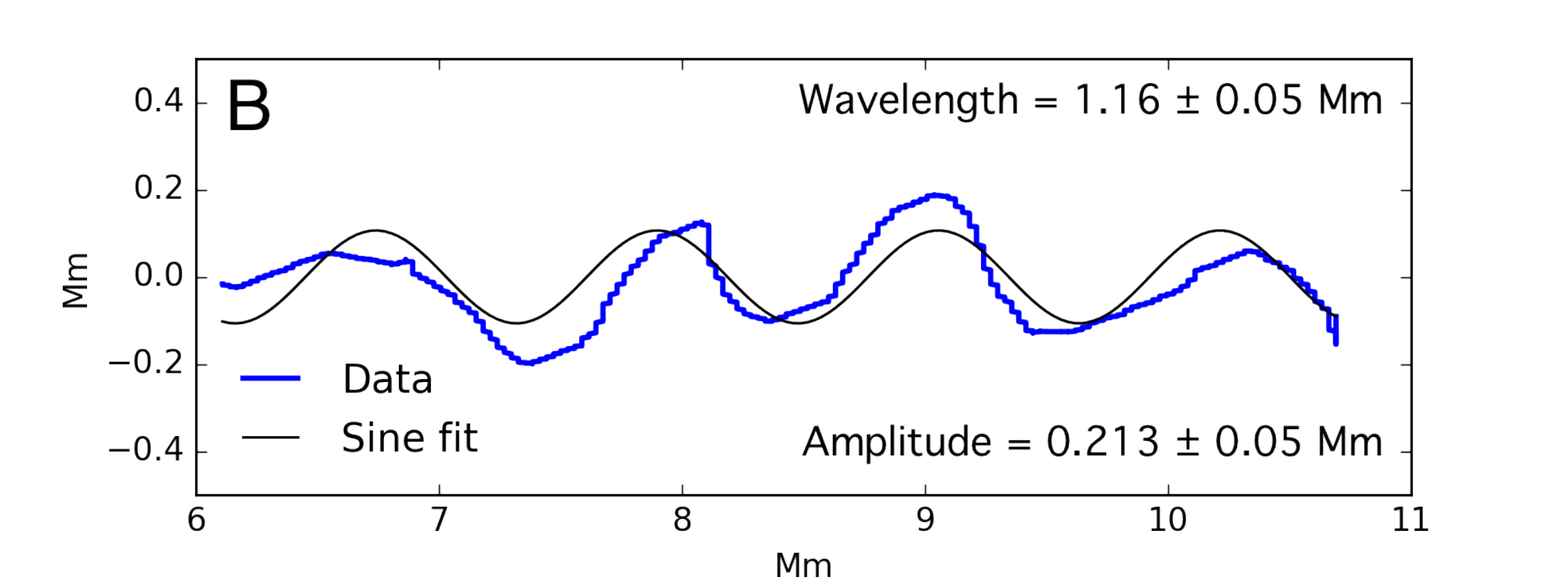}
\includegraphics[width=0.485\textwidth,trim=0mm 0mm 4mm 0mm, clip]{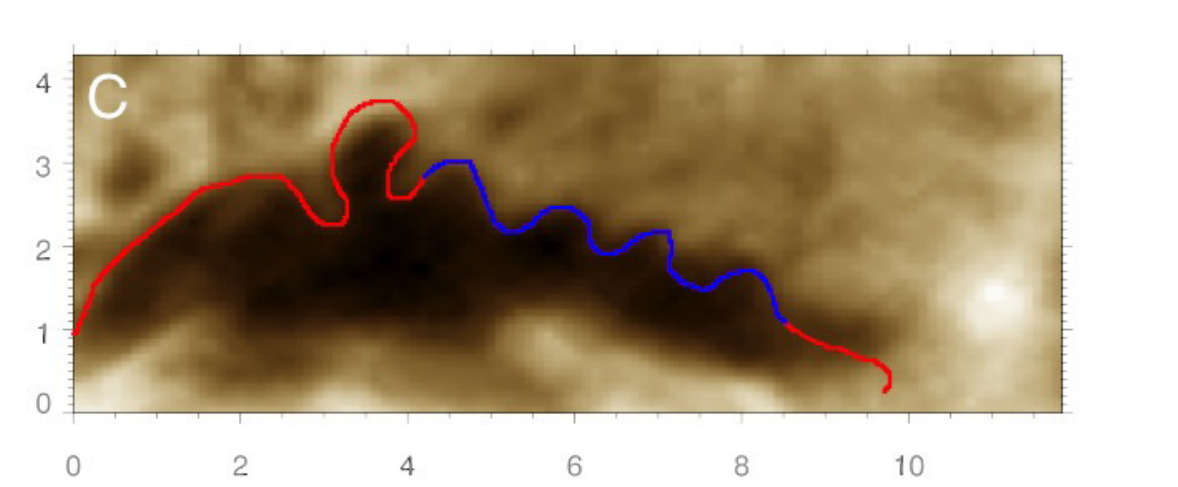}
\includegraphics[width=0.495\textwidth]{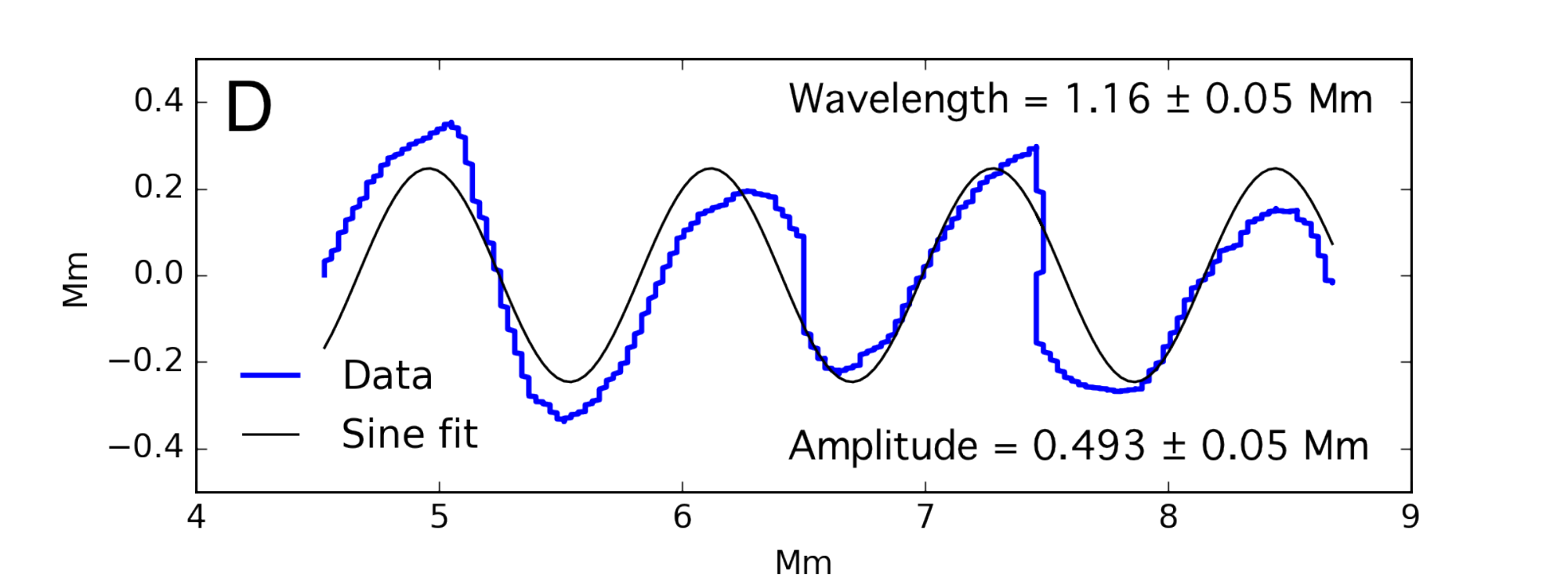}
\caption{Closeup of the 08-August-2007 bubble boundary prior to plume formation. (A) Enlarged image of the prominence bubble from Frame~699 at 20:00:42~UT with manually defined boundary plotted in color. The blue section of the boundary is shown in Panel~B. Tickmarks are Mm. (B)  Blue boundary section from Panel~A with a least-squares linear fit subtracted. Least-squares sinusoidal fit to the boundary perturbations is plotted in black. (C) and (D): same as (A) and (B) but for Frame 709 at 20:02:24~UT. Note that the zero-point of the y-axis is not the same as in Fig.~\ref{fig:0808_montage}. \revision{Note that error levels in wavelength and amplitude measurements are generated from enlarged and interpolated images used to manually define the profiles and thus represent sub-pixel centroid accuracies}. }
\end{center}
\label{fig:0808_bumps}
\end{figure}

In Frame~690 (19:59:13~UT), the initial flowfield in the region of the bubble is downward and somewhat perpendicular to the bubble boundary, with a typical average prominence downflow speed of about 9$\kmps$. There is evidence of more horizontal flow at the right edge of the bubble, but at a significantly lower speed. However about 1.5~min later, in Frame~699 (20:00:42~UT), the flow just above the bubble boundary has rotated to be more aligned to the boundary. Simultaneously, the bubble boundary develops a regular perturbation, or ``bumps,'' that are shown in detail in Fig.~\ref{fig:0808_bumps}A. 

The perturbations identified in Frame~699 propagate along the boundary while the plasma above the bubble begins to rotate and follow the boundary flow. By Frame~703 (20:01:22~UT) the flow has begun to acquire a significant upward component near the top of the bubble. In Frame~709 (20:02:24~UT), plume formation has initiated. The total time from initial boundary shear flow alignment and perturbation to plume onset is approximately 2~min. The white lines in the lower panels of Fig.~~\ref{fig:0808_montage} demarcate the top of the first major plume and show that it achieves a scale of about 2~Mm across in 96~s while rising with a speed of 21~$\kmps$. In Frame~719 (20:04:04) there is evidence of plasma intrusion across the boundary from the right side as the bubble collapses and the flow becomes multi-directional in the region of the bubble. 

Figure~\ref{fig:0808_bumps}A shows an enlargement of the bubble in Frame 699 of Fig.~\ref{fig:0808_montage} with the boundary pixels defined manually using an interactive display tool. The blue section of the boundaries in Fig.~\ref{fig:0808_bumps} highlights the regular ``bumps'' mentioned above in order to analyze their growth over time.  We interpret these bumps as an indication of KHI onset, caused by the strong shear flow along the boundary between the bubble and the overlying prominence. The unique feature of this instability however is that it is taking place on the boundary of a density inversion, presenting the possibility of a hybrid KH-RT instability. 

Figure~\ref{fig:0808_bumps}B shows the boundary perturbations enlarged further and normalized by a least-squares linear fit. A least-squares curve fitting procedure is used to define the best fit sine curve in order to measure the amplitude and wavelength of the perturbations.  The sine fit to Frame~699 gives a wavelength $\lambda = 1.16\pm0.05$~Mm and a peak-to-peak amplitude  $a = 0.213\pm0.05$~Mm (213~km). Figures~\ref{fig:0808_bumps}C and D repeat the analysis for Frame 709 from Fig.~\ref{fig:0808_montage} (taken 162~s after Frame~699) and show an amplitude of 0.493$\pm0.05$~Mm (493~km) while the wavelength remains essentially unchanged. The relatively stable wavelength of the growing perturbations is characteristic of a hybrid KH-RT perturbation mode developing over time. 

The exponential growth rate of this mode can be \revision{approximated by}
\begin{equation}
	\gamma_{\rm OBS} = \frac{1}{(t-t_0)} \ln\left(\frac{h - h_0}{h_0} + 1\right)
\label{eqn:gr}
\end{equation}
where the overbar indicates that this growth rate is an approximate value based on measurements. With $h - h_0 = 0.280\pm0.12$~Mm and $t - t_0 = 162$~s we find $\gamma_{\rm OBS} = 0.0052\pm0.0015\mathrm{s}^{-1}$. 

The perturbation structures on the bubble boundary are not static; they move along the boundary with a characteristic phase speed that is a function of the shear flow magnitude and the plasma and magnetic field conditions in the prominence and in the bubble. Using a series of enlarged images of the prominence bubble boundary and manually tracking four separate structures with an estimated error in position of four pixels, phase velocities of between $17.7\pm0.2$ and $26.0\pm0.2\kmps$ were measured. The mean phase velocity for the perturbations is approximately $22\kmps$.

\section{Coupled Kelvin-Helmholtz -- Rayleigh-Taylor Instability Analysis}
\label{sec:khrt}

In this section we assume that the perturbations measured above are the initial stages of a hybrid KH-RT instability and use analytical MHD formulations of this hybrid instability to infer velocity and magnetic field characteristics of the system. We note that lateral deflection of prominence downflows at bubble boundaries is a common feature of SOT and ground-based prominence bubble observations. Thus while not all bubble boundaries form well-defined shear flow structures such as those analyzed in Sec.~\ref{sec:shear}, we believe that the stability of all prominence bubbles is likely influenced by shear flows to some extent. We can deduce properties of the magnetic field in the prominence bubble system by comparing measured and estimated quantities with analytic expressions derived from linear perturbations to the ideal MHD equations \citep[e.g.,][]{Zhang:2005it}. 

Specifically, the characteristics of the shear flow structures measured in Sec.~\ref{sec:shear} \revision{such as the phase velocity of the perturbations and the measured growth rate of the instability prior to the non-linear plume formation phase} can be compared to analytical expressions for the hybrid MHD KH-RT \revision{instability}. 
\revision{Analyzing perturbation waves on the boundary between two layers with a horizontal interface where the layers have different densities (the heavier being above the lighter) and velocities, combined with the assumption of incompressibility and a uniform horizontal magnetic field at an arbitrary direction to the flow, the frequency for a perturbation wave vector $\mathbf{k}$ with magnitude $k = 2\pi/\lambda$ is given by (see Appendix for derivation):}
\begin{equation}
	\omega = (\alpha_+\mathbf{k\cdot V}_+ +\alpha_- \mathbf{k\cdot V}_-) \pm i \sqrt{gkA-\frac{(\mathbf{k\cdot B})^2}{2\pi(\rho_++\rho_-)}+\alpha_+\alpha_-(\mathbf{k\cdot \Delta V})^2}				
\label{eqn:fullomega}
\end{equation}
where $\rho_\pm$ are the prominence and bubble densities, respectively, $\alpha_{\pm} = \rho_{\pm}/(\rho_{+} + \rho_{-})$ is a dimensionless density ratio,  $\mathbf{V}_{\pm}$ are the plasma flow vectors above and below the density inversion boundary, respectively,  $g$ is the acceleration of gravity, $A$ is the Atwood number, $\mathbf{B}$ is the local magnetic field vector, $\mu_{0}$ is the permeability of free space, and $\mathbf{\Delta V = V_{+} - V_{-}}$ is the shear flow velocity vector along the boundary between the layers. 

\revision{When substituted into the perturbation equation time dependency given by $\exp(-i\omega t)$,  $\mathrm{Re}(\omega) \equiv \omega_{\rm R}$ results in an oscillatory solution that defines the phase velocity of the perturbations moving along the shear flow boundary; similarly, the negative term of $\mathrm{Im}(\omega)$ results in an exponentially growing component that defines the linear phase of the KH and RT instability flows. Hereafter we use the common notation $\mathrm{Im}(\omega) \equiv \gamma$ for the growth rate and compare it to $\gamma_{\rm OBS}$ from Equation~(\ref{eqn:gr}). }

\revision{We note that the expression for $\gamma$ assumes an equal magnetic flux density in the prominence and the bubble, and a negligibly thin boundary between the density regions. Also, if the term inside the square-root becomes negative due to, e.g., a strong enough magnetic field component along the perturbation wave vector, then $\mathrm{Im}(\omega) = 0$ and a pure surface wave solution is obtained.}

Substituting the expression $\mathbf{V_{-} = V_{+} - \Delta V}$ into $\omega_{\rm R}$ reveals an expression for the phase velocity that we can solve for $\Delta V$: 
\begin{equation}
	v_{p} = \frac{\omega_{\rm R}}{k} = \frac{1}{k}\left(\mathbf{k}\cdot\mathbf{V_{+}} - \alpha_{-}\mathbf{k}\cdot\mathbf{\Delta V}\right)
\label{eqn:vphase}
\end{equation}
where $v_{p}$ is the phase velocity. This gives an equation for the velocity shear in terms of the measured phase velocity, the observed wave vector, and the density ratio $\alpha_{-}$: 
\begin{equation}
	\mathbf{k\cdot\Delta V} = \frac{1}{\alpha_{-}}\left(\mathbf{k\cdot V_{+}} - kv_{p}\right)
\label{eqn:kdeltau}
\end{equation}

Examining the expression for the \revision{instability} growth rate given by $\mathrm{Im}(\omega) \equiv \gamma$ in Eqn.~\ref{eqn:fullomega}, we see that the first two terms are the standard magnetic RTI terms \citep[e.g.,][]{Chandrasekhar:1961a} and the third term describes the influence on the growth rate of the shear flow across the density inversion.  Grouping terms differently gives
\begin{equation}
	\gamma^2 = gkA - \left(\frac{2 (\mathbf{k}\cdot\mathbf{B})^{2}}{\mu_{0}(\rho_{+}+\rho_{-})} - \alpha_{+}\alpha_{-}(\mathbf{k\cdot\Delta V})^2\right)
\label{eqn:gammai}
\end{equation}
The first term in the expression is the hydrodynamic RTI growth rate. The term in parentheses can be viewed as a balance between the magnetic tension force across the density inversion boundary and the shear flow, showing that shear flow works to decrease the tension across the boundary and hence increase the growth rate of the hybrid KH-RT instability.  However the addition of shear flow does not modify the fact if any boundary perturbation has a wave vector component perpendicular to the magnetic field (i.e., $\mathbf{k}\cdot\mathbf{B} = 0$), then that component will always be unstable. The implication is that in a real 3D physical system with a multitude of perturbations, some of which are bound to be perpendicular to the local magnetic field, a uniform magnetic field can never suppress the RTI \citep{Stone:2007jx}; adding shear flow across the boundary only increases the instability of the system. \cite{Ruderman:2014cq} show that a nonuniform vertically sheared magnetic field decreases the growth rate of the RTI relative to the uniform field case, but does not fully suppress it. 

Using Eqns.~\ref{eqn:kdeltau} and \ref{eqn:gammai} to calculate shear flow and magnetic field characteristics requires making assumptions about the prominence and bubble mass densities. Since prominence density measurements in the literature are sensitive to both the type of prominence and the location in the prominence where the spectroscopic measurements are taken, we avoid assigning a single number and instead assume a range of values based on the recent review of \cite{Labrosse:2010bt}. Thus we take \revision{a prominence density} $\rho_+ = 10^{-11}$ to $10^{-9}\kgpcm$ ($10^{-14}$---$10^{-12}\gpcc$) and use bubble densities\footnote{\revision{Measurements of prominence bubble thermodynamic properties have not yet been definitive. Since the bubbles are dark in \CaIIH and \Ha lines, their density is either much lower than typical prominence density values, and/or their internal temperatures are high enough to ionize Hydrogen. \cite{Berger:2011eoa} find evidence for bubble internal temperatures of at least 250,000K but these measurements have not been confirmed. For the purposes of this analysis we assume density values in the bubbles that are one order-of-magnitude below typical quiescent prominence density values, thus making the bubble densities commensurate with typical coronal density values.}} $\rho_-$ in the range $10^{-12}$ to $10^{-11}\kgpcm$. We also note that quiescent prominences, particularly the ``hedgerow'' variety like the one analyzed in Sec.~\ref{sec:shear}, are generally thin, sheet-like structures. In this case, we can reasonably assume that \revision{the} ``plane'' of the prominence is roughly in the plane of the sky \revision{and} that the line-of-sight is perpendicular to the prominence plane. Thus there are no projection distortions and the value of $\lambda$ that we measure in Sec.~\ref{sec:shear} is the wavelength of the wave vector in the plane of the prominence, i.e., the longitudinal component along the prominence plane  $\lambda_{l} = 1.16$~Mm, which implies $|\mathbf{k_l}| = 5.42\times 10^{-6}\ \mathrm{m}^{-1}$. Finally, we use an approximate value of $V_+ \sim 10\kmps$ from the NAVE flow tracking speeds in the shear flow zone, the mean phase speed value of $v_p = 22\kmps$, and a solar gravity of $274\mpss$. 

\begin{figure}
\begin{center}
\includegraphics[width=0.435\textwidth]{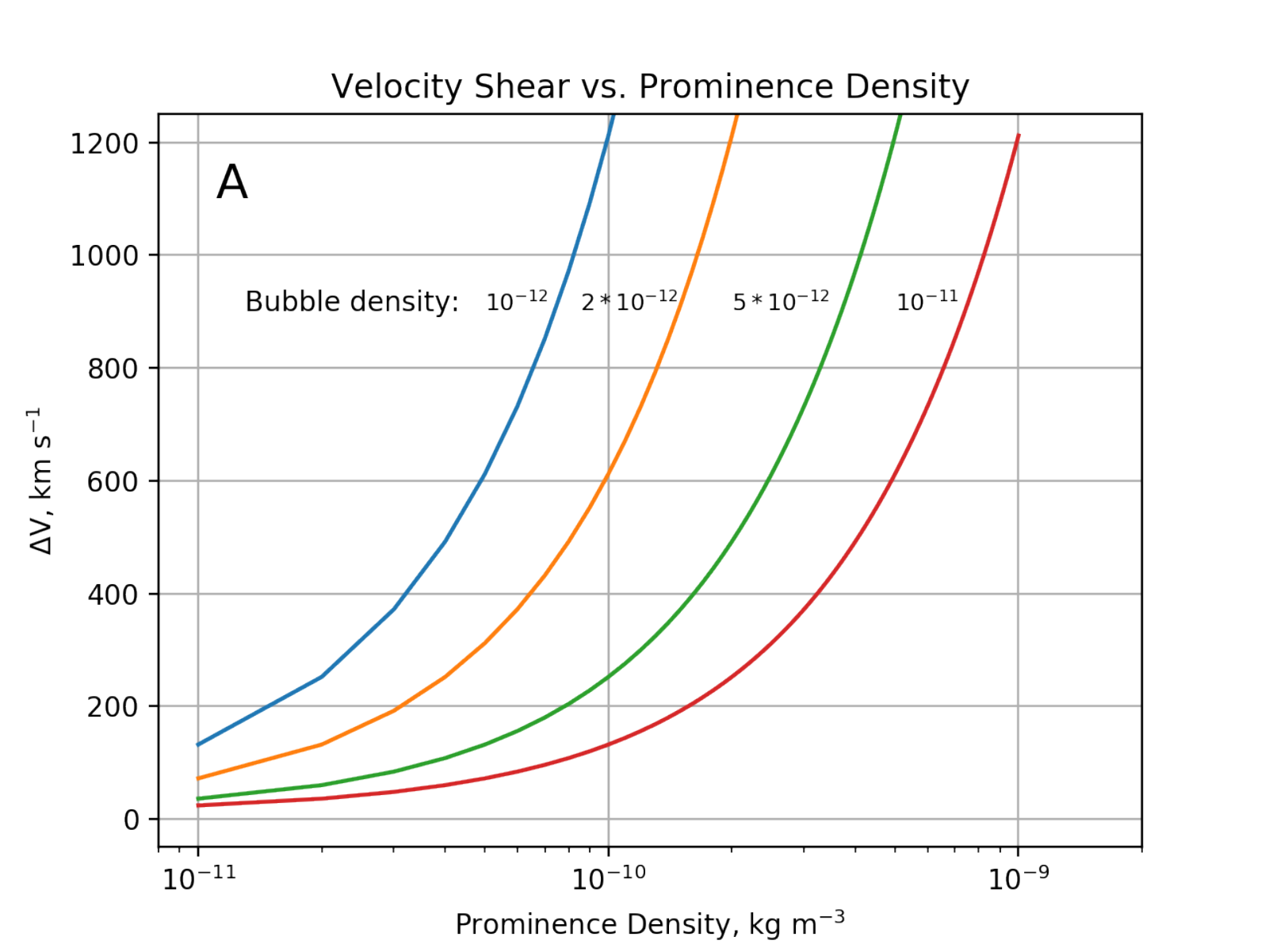}
\includegraphics[width=0.435\textwidth]{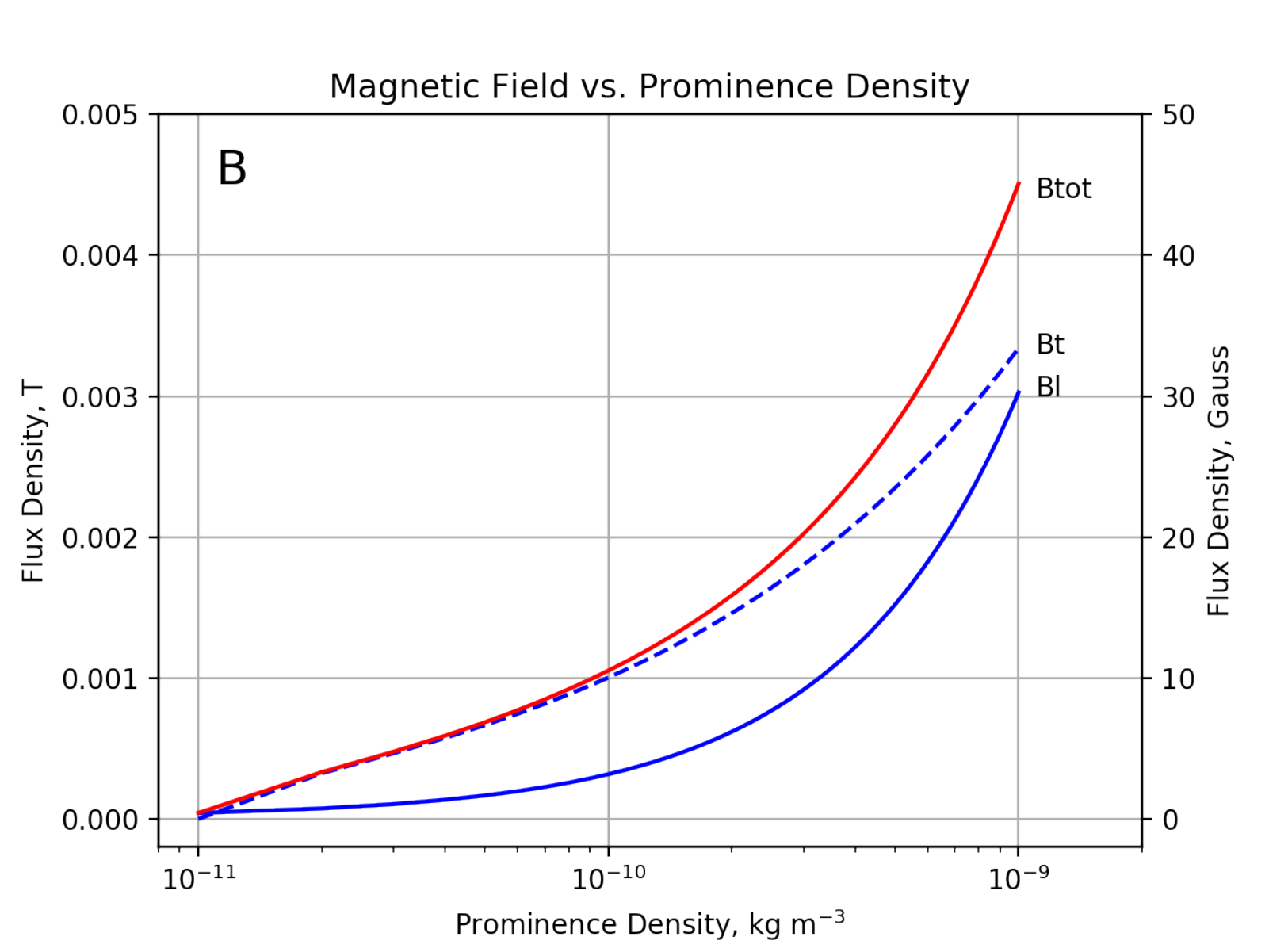}
\caption{Velocity shear across the prominence bubble (A) and inferred magnetic flux density (B) as a function of prominence mass density from a magnetic KH-RT instability analysis using the values of $V_+$, $v_p$, and perturbation wavelength $\lambda_l$ measured in Sec.~\ref{sec:shear}. In Panel A, the bubble density ranges from $10^{-12}$ to $10^{-11}\kgpcm$. In Panel B, the bubble density is set to $10^{-11}\kgpcm$ since for typical prominence densities in the range of $10^{-10}$ to $10^{-9}\kgpcm$ this gives flux densities consistent with current measurements. In Panel B, $\mathrm{B_l}$ is the longitudinal magnetic field in the plane of the prominence, $\mathrm{B_t}$ is the transverse field, and $\mathrm{B_{tot}}$ is the total flux density of the field. }
\end{center}
\label{fig:deltav}
\end{figure}

For illustration, Fig.~\ref{fig:deltav}A plots the shear flow magnitude, $\Delta V$ against prominence density for a range of bubble densities that vary $\alpha_-$ in Eqn.~\ref{eqn:kdeltau}. As expected, as the prominence density approaches the bubble density value, the shear flow required to drive the observed phase speed of the perturbation also decreases. Taking a canonical prominence density value of $10^{-10}\kgpcm$, and a bubble density of $10^{-11}\kgpcm$, $\Delta V\sim130\kmps$ which implies a flow velocity in the bubble of $V_- = 120$---$140\kmps$ depending on whether it is antiparallel or parallel to $V_+$. Taking the average value and equating it to the Alfv{\'e}n speed in the bubble, i.e. assuming that the flow in the bubble is driven primarily by Lorentz forces during the flux emergence process, the estimated magnetic field in the bubble is $5\times10^{-4}$~T (5~gauss). 

Again using prominence and bubble density values of $10^{-10}$ and $10^{-11}\kgpcm$, respectively, the Atwood number $A=0.82$ and the static hydrodynamic RTI growth rate for a perturbation with $k=k_l$ is given by $\gamma_{\mathrm{hydro}} = \sqrt{gk_lA} = 0.0349\ps$. The shear flow term for this case has a magnitude of $\gamma_{\mathrm{shear}} = 0.202\ps$ implying that the total expected growth rate for the non-magnetic case is $\gamma_{0} = 0.2055\ps$. But the measured growth rate of $0.0052\  \mathrm{s}^{-1}$ is far less than this value, implying that the magnetic field is suppressing the growth rate significantly through tension applied along the direction of $\mathbf{k_l}$.  To estimate the magnetic field strength, we can approximate a decomposition of the magnetic field term in Eqn.~\ref{eqn:gammai} into longitudinal and transverse components that are in, and perpendicular to, the prominence plane, respectively. For the transverse component, we assume a perturbation wavevector $k_t$ that propagates perpendicular to the prominence sheet:
\begin{equation}
	\gamma^2 = gk_lA + \alpha_{+}\alpha_{-}k_{l}^2\Delta V^2  - \frac{(k_l^2B_l^2 + k_t^2B_t^2)}{\mu_{0}\bar\rho}
\label{eqn:kl}
\end{equation}
where $\bar\rho$ is the average of the upper and lower density values and we have approximated $k_l \gg k_t$ for this system. Taking the derivative of this expression with respect to $k_l$ and equating the result to zero gives the expression for the most unstable longitudinal mode, which we assume is the perturbation measured in Sec.~\ref{sec:shear}. This gives
\begin{equation}
	B_l^2 = \mu_0\bar\rho\left(\frac{gA}{2k_l} + \alpha_+\alpha_-\Delta V^2\right)
\end{equation}
Finally, to determine the value of $B_t$ we use the measured growth rate of the longitudinal perturbation, $\gamma_{OBS} = 0.0052\ps$, and rearrange Eqn.~\ref{eqn:kl} to get 
\begin{equation}
	B_t^2 =  \frac{\mu_0\bar\rho}{k_t^2} \left( gk_lA   + \alpha_+\alpha_-k_l^2\Delta V^2 - \gamma^2\right) - \frac{k_l^2B_l^2}{k_t^2}
\end{equation}
Assuming that $k_t$ is determined by the length scale of a hemispherical bubble, i.e., $\lambda_t \sim 10$~Mm, $k_t \sim 2\times10^{-7}\ \mathrm{m}^{-1}$, Fig.~\ref{fig:deltav}B plots the range of magnetic flux density as a function of prominence density for a constant bubble density of $10^{-11}\kgpcm$. For the canonical prominence density of $10^{-10}\kgpcm$ this results in longitudinal and transverse flux density values of $B_l = 3.2\times10^{-4}$~T (3.2~gauss) and $B_t = 1.0\times10^{-3}$~T (10~gauss). The total flux density across the prominence is approximately equal to $B_t$ in this case, or about 10~gauss, in general agreement with the value derived from the shear flow analysis above. The angle of the field with respect to the prominence normal is $\arctan(3.2/10)\sim 18\degree$, implying that the field is nearly perpendicular to the prominence plane. We note that using a prominence density of $10^{-9}\kgpcm$, two orders of magnitude above the bubble density, implies total flux density values of 40---50~gauss but with the field vector more aligned with the prominence axis. This range of flux density values and angles is within the range of measured vector magnetic fields in prominences \citep[see, e.g.,][]{OrozcoSuarez:2014he}. More to the point, this analysis shows that interpreting the perturbation and instability measured in Sec.~\ref{sec:shear} as a hybrid KH-RT instability results in inferred values for the magnetic field that are consistent with generally accepted properties of quiescent polar crown prominences.   

\section{Summary and Conclusions}
\label{sec:summary}

The analyses of the previous sections have established the following:
\begin{enumerate}
\item Prominence bubbles can experience steady growth before instability onset, with characteristics rise speeds of a few $\kmps$. Instability onset increases these speeds up to characteristic values on the order of $10\kmps$. 

\item Bubble boundary layer brightness varies gradually as a function of time. Characteristic thickness of the brightest plasma in the layer varies between 0.25 and 0.5~Mm and is correlated with the gradual buildup of plasma flowing down from the prominence above. We see no evidence of \revision{localized} impulsive brightening events on bubble boundaries. 

\item ``Drainage downflows'' can generate from thickened bubble boundaries and achieve characteristic speeds of 25--$50\kmps$. These flows can follow apparently arched trajectories around the bubble, i.e., they do not necessarily fall vertically from the boundary. 

\item Downflows of prominence plasma onto the boundary layer can be redirected laterally to flow along the boundary. These flows, and presumably flows interior to the bubble, can establish strong shear flows across the boundary that lead to characteristic KH vortex formation. The measured phase velocity of the bubble analyzed in Sec.~\ref{sec:shear} implies a shear flow magnitude on the order of $100 \kmps$ across the bubble boundary.

\item Bubble boundary shear flows can result in the sudden destabilization of the bubble  as seen in Sec.~\ref{sec:growth}, or in the development of regular structures resembling KH flows that couple with the RTI to modify the scale of the plume upflows, as seen in Sec.~\ref{sec:shear}. 

\item Coupled KH-RT instability analysis of the 08-August-2007 prominence bubble event \revision{assuming typical density values} implies a strongly sheared prominence magnetic field on the order of $10^{-3}$~T (10~gauss) at an angle of about $70\degree$ to the prominence plane. The sheared magnetic field compensates for the strong velocity shear across the bubble boundary layer to significantly suppress the growth rate of the RTI in this prominence.  

\end{enumerate}

Items 1--3 are consistent with the hypothesis that prominence bubbles are formed by an emerging magnetic flux complex below a pre-existing prominence. The thermodynamic and magnetic pressure within the emerging flux complex lead to buoyancy forces that cause the complex to expand and slowly rise into the overlying chromosphere and prominence until magnetic and thermodynamic pressures are in equilibrium with the surroundings. The ubiquitous quiescent prominence downflows accumulate in a boundary layer that appears as an enhanced brightness layer in both the Ca~II and H$\alpha$ bandpasses. Boundary \revision{layer brightness and thickness increase gradually on time scales of minutes in wide regions that are correlated with downflows of prominence plasma}. When the built-up boundary layer is no longer gravitationally stable, downflows initiate that drain the layer and decrease its thickness and brightness. 

 \revision{The observations analyzed here do not seem to support a magnetic reconnection mechanism for boundary brightening or downflow initiation since we find no evidence for small-scale, impulsive (i.e. on time scales of the Alfv{\'e}n transit time across the layer which, for a thickness on the order of 0.5~Mm and $v_{A} = 30 \kmps$ is approximately 15~s), bright spot formation in the movies of these events. In addition, reconnection would be expected to produce bi-directional flows from the reconnection site.  Although we find no evidence of coincidence between drainage downflows and impulsive upflows that might indicate bi-directional reconnection outflows from the boundary layer, reconnection outflows may be difficult to detect and cannot necessarily be ruled out by these results. However, the observed downflows are complex in structure and orientation and appear to have velocity components out of the plane of the sky, i.e. the flows seem more consistent with plasma falling under the influence of gravity around a hemispherical magnetic domain than with reconnection outflows.}

Items 4--6 establish that prominence bubbles offer a rich environment for studies of MHD flow instabilities in the solar atmosphere. The analyses above show that in addition to the large-scale bubble inflation and RTI plume dynamics, small-scale shear flows on the bubble boundary offer yet another avenue for detailed study of magnetic plasma instabilities. We conclude that solar quiescent prominence bubble boundary dynamics are fully consistent with the characteristics predicted by theoretical and numerical studies of the coupled Kelvin-Helmholtz and Rayleigh-Taylor instabilities.   For the case analyzed in Sec.~\ref{sec:shear}, a magnetic field of approximately 10~gauss at an angle of about $70\degree$ to the prominence sheet is enough to suppress the instability growth rate to about 15\% of the hydrodynamic ($B=0$) value. We note that by the time the perturbation on the boundary is visible in the SOT data, it is already nearing the non-linear evolution phase and the structures we observe as ``bumps'' are likely well-developed vortex flows moving at the observed phase speed along the boundary. This argues for a higher resolution solar optical telescope that can directly observe the earlier linear stages of these MHD instabilities. 

In summary, we have analyzed the first observations, to our knowledge, of the coupled KH-RT instability in an astrophysical system. Additional analyses of prominence bubble instabilities using both SOT data and new observations from the NASA Interface Region Imaging Spectrometer mission \citep[IRIS,][]{DePontieu:2014aa} are ongoing. The addition of the IRIS spectral observations will enable doppler velocity measurements to remove the ambiguities surrounding the directions of the downflows, plumes, and boundary shear flows relative to the line-of-sight. The addition of shear flows to existing ideal and partially ionized MHD models of prominence bubbles could yield new insights into the reduced growth rates that we measure compared to the standard RTI analyses completed to date. Understanding the detailed flow dynamics of phenomena such as the coupled Kelvin-Helmholtz and Rayleigh-Taylor instabilities will aid our understanding of energy transfer and flow dynamics in other solar phenomena such as magnetic eruptions \citep[e.g.,][]{Foullon:2011cw,Ofman:2011gs,Mostl:2013ir}, and could lead to insights into prominence eruptions as well. 

Determination of whether prominence bubbles are complex, 3D, dome-like structures created by emerging flux heated to coronal temperatures, or simply arches formed by ``parasitic bipole'' perturbations to the existing prominence field structure will require high-resolution optical and EUV observations of bubble events in filaments on the disk. These events would presumably be visible as rapid disruptions of the filament structure as the instability upflows passed through.  Along with simultaneous magnetogram measurements to determine the type and location of flux emergence that accompanies these events, DEM analysis of the plasma associated with the flux emergence, free from any foreground or background confusion, would determine whether there was a thermal component helping to drive an RTI instability. 

Finally, we note that analyses of prominence dynamics such as those performed here were impossible before the flight of the \emph{Hinode} satellite. \emph{Hinode}/SOT has demonstrated the discovery potential of a medium resolution solar optical telescope in orbit by enabling uninterrupted diffraction-limited prominence observations with durations measured in hours. In contrast, diffraction-limited observations of prominences from ground-based telescopes have typical durations measured in minutes due to the challenges of atmospheric seeing. It is perhaps not surprising that in the 75 years of prominence literature prior to the \emph{Hinode} mission there were only two published observations of prominence bubble events \citep{Stellmacher:1973vg,deToma:2008ei} \revision{with analyses that were severely limited by low spatial and temporal resolution}. While the SOT data enable the detailed analyses above, \revision{further improvements in spatial and temporal resolution are required in order} to study the small-scale details of prominence plasma instability onset as well as the details of the flows in the non-linear stage. With the upcoming implementation of advanced adaptive optics on the ground-based 4-meter Daniel K. Inouye Solar Telescope (DKIST), it will be interesting to see if ground-based prominence observations can achieve long enough durations to enable these analyses.  Even if successful, ground-based telescopes will always be limited by daylight time and weather and will never surpass orbital platforms in terms of continuous targeting and duration of diffraction-limited observations. Given the recent failure of the SOT filtergram camera, it is imperative that we begin planning for an innovative space-based optical solar telescope that can follow on the SOT discoveries by delivering high-resolution Hanle and Zeeman effect magnetic field measurements in prominences over the hours-long evolutionary periods of these enigmatic solar structures.  

\section*{Appendix: Derivation of Eqn. 2, the KHRT Instability Growth Rate}

Assuming two volumes of constant density, with the larger density above the smaller, in a plane-parallel atmosphere with a magnetic field that is purely horizontal, the initial conditions for the magnetic Kelvin-Helmholtz-Rayleigh-Taylor instability are given by:
\begin{align}
\rho_0=&\begin{cases} \rho_+ & \mbox{if $z > 0$};\\ \rho_- & \mbox{if $z < 0$}\end{cases} \\
V_{y0}=&\begin{cases} V_{y_+} & \mbox{if $z > 0$};\\ V_{y_-} & \mbox{if $z < 0$}\end{cases} \\
V_{x0}=&\begin{cases} V_{x_+} & \mbox{if $z > 0$};\\ V_{x_-} & \mbox{if $z < 0$}\end{cases} \\
V_{z0}=&\ 0 \\ 
P=&\ P_0-g\rho_0 z\\
B_{x0}=&\ B_x\\
B_{y0}=&\ B_y\\
B_{z0}=&\ 0.
\end{align}

The linearized MHD equations in uniform gravity are:
\begin{align}
\frac{\partial \rho'}{\partial t}+\mathbf{V}_0\cdot \nabla\rho'+v_z\frac{\partial \rho_0}{\partial z}=0\\
\rho_0\left[\frac{\partial}{\partial t} + \mathbf{V}_0\cdot \nabla \right]\mathbf{v}=-\partial p+\mathbf{j}\times\mathbf{B}+\rho'\mathbf{g}\\
\left[\frac{\partial}{\partial t} + \mathbf{V}_0\cdot \nabla \right]\mathbf{b}=\mathbf{B}\cdot\nabla \mathbf{v}\\
\nabla\cdot\mathbf{v}=\nabla\cdot\mathbf{b}=0
\end{align}
where $\mathbf{v}$, $\mathbf{b}$, $p$ and $\rho'$ are the perturbed quantities from the initial equilibrium, i.e., $\bf V = V_0 + v$, $\rho = \rho_0 + \rho'$, etc., with $\bf v << V_0$, $\rho' << \rho_0$, etc.
Using a normal mode decomposition in the horizontal plane we look at perturbations to this system of the form $f(x,y,z,t)=\tilde{f}(z)exp(ik_xx+ik_yy-i\omega t)$. 
By manipulating these equations to get a single ODE in terms of $v_z$, where a consequence of this is that we find $ik_yv_x-ik_xv_y=ik_yb_x-ik_xb_y=0$, we ultimately get:
\begin{equation}\label{v_z_eqn}
\rho_0\Omega^2k^2v_z=\Omega d_z\left(\rho_0\Omega d_zv_z\right) +k^2\frac{(\mathbf{k\cdot B})^2}{4\pi}v_z-\Omega\frac{(\mathbf{k\cdot B})^2}{4\pi}d_z\left(\frac{1}{\Omega}d_z v_z\right)-k^2g(d_z\rho_0)v_z
\end{equation}
where $\Omega=\mathbf{k\cdot V_0} -\omega$ and $d_z$ is the $z$ derivative.
For $z\ne0$ the solution of $\tilde{v}(z)$ is $\tilde{v}(z)=exp(-k|z|)$ (as this guarantees that $\tilde{v}(z)=0$ at $z=\pm\infty$).

Integrating Equation (\ref{v_z_eqn}) across the contact discontinuity leads to:
\begin{equation}
\rho_+(k_xV_{x_+} + k_yV_{y_+} - \omega)^2+\rho_-(k_xV_{x_-} + k_yV_{y_-} - \omega)^2=\frac{(\mathbf{k\cdot B})^2}{2\pi}-(\rho_+-\rho_-)gk.
\end{equation}
Then on expanding and simplifying, the solution for $\omega$ is:
\begin{equation}
\omega = (\alpha_+\mathbf{k\cdot V}_+ +\alpha_- \mathbf{k\cdot V}_-) \pm i \sqrt{gkA-\frac{(\mathbf{k\cdot B})^2}{2\pi(\rho_++\rho_-)}+\alpha_+\alpha_-(\mathbf{k\cdot \Delta V})^2},
\end{equation}
where $\alpha_{+/-}=\rho_{+/-}/(\rho_++\rho_-)$ and the Atwood number $A=\alpha_+-\alpha_-$. This is the same as Eqn.~\ref{eqn:fullomega} in Sec.~\ref{sec:khrt}.


\acknowledgments
\revision{We thank Professor Jongchul Chae of the Department of Physics and Astronomy, Seoul National University, Korea, for supplying the NAVE optical flow tracking code used to analyze the 08-August-2007 prominence, Dr. Craig DeForest of SWRI, Boulder, for supplying the SOT/BFI instrument PSF kernel used to process all prominence images analyzed here, and the anonymous referee for helpful suggestions on improving the manuscript.  \emph{Hinode} is a Japanese mission developed and launched by ISAS/JAXA, with NAOJ as domestic partner and NASA and STFC (UK) as international partners. It is operated by these agencies in cooperation with ESA and NSC (Norway). T.E.B. was supported by NASA contracts NNM07AA01C (Solar-B FPP), NNG04EA00C (SDO/AIA) while at the Lockheed Martin Solar and Astrophysics Laboratory (LMSAL), and by The National Weather Service (NWS) Office of Science and Technology Integration (OSTI) while at the National Oceanic and Atmospheric Administration (NOAA). A.H was supported by his STFC Ernest Rutherford Fellowship grant number ST/L00397X/2. W.L. was supported by NASA HGI grant NNX15AR15G and NASA contract NNG09FA40C (IRIS) at LMSAL.  Data analysis was coded in both IDL and the NumPy module of Python.}


\begin{thebibliography}{}

\bibitem[\protect\astroncite{Antolin and Rouppe van~der
  Voort}{2012}]{Antolin:2012jz}
Antolin, P. and Rouppe van~der Voort, L.: 2012,
\newblock {\em ApJ} {\bf 745(2)}, 152

\bibitem[\protect\astroncite{Aulanier and Demoulin}{1998}]{Aulanier:1998tn}
Aulanier, G. and Demoulin, P.: 1998,
\newblock {\em A{\&}A} {\bf 329}, 1125

\bibitem[\protect\astroncite{Bak-Ste{\'{s}}licka
  et~al.}{2013}]{BakSteslicka:2013hg}
Bak-Ste{\'{s}}licka, U., Gibson, S.~E., Fan, Y., Bethge, C., Forland, B., and
  Rachmeler, L.~A.: 2013,
\newblock {\em ApJ} {\bf 770(2)}, L28

\bibitem[\protect\astroncite{Berger et~al.}{2011}]{Berger:2011eoa}
Berger, T., Testa, P., Hillier, A., Boerner, P., Low, B.~C., Shibata, K.,
  Schrijver, C., Tarbell, T., and Title, A.: 2011,
\newblock {\em Nature} {\bf 472(7342)}, 197

\bibitem[\protect\astroncite{Berger et~al.}{2012}]{Berger:2012da}
Berger, T.~E., Liu, W., and Low, B.~C.: 2012,
\newblock {\em ApJ} {\bf 758(2)}, L37

\bibitem[\protect\astroncite{Berger et~al.}{2008}]{Berger:2008kx}
Berger, T.~E., Shine, R.~A., Slater, G.~L., Tarbell, T.~D., Title, A.~M.,
  Okamoto, T.~J., Ichimoto, K., Katsukawa, Y., Suematsu, Y., Tsuneta, S.,
  Lites, B.~W., and Shimizu, T.: 2008,
\newblock {\em ApJ} {\bf 676(1)}, L89

\bibitem[\protect\astroncite{Berger et~al.}{2010}]{Berger:2010bj}
Berger, T.~E., Slater, G., Hurlburt, N., Shine, R., Tarbell, T., Title, A.,
  Lites, B.~W., Okamoto, T.~J., Ichimoto, K., Katsukawa, Y., Magara, T.,
  Suematsu, Y., and Shimizu, T.: 2010,
\newblock {\em Astrophysical Journal} {\bf 716(2)}, 1288

\bibitem[\protect\astroncite{Chae}{2010}]{Chae:2010ds}
Chae, J.: 2010,
\newblock {\em ApJ} {\bf 714(1)}, 618

\bibitem[\protect\astroncite{Chae et~al.}{2008}]{Chae:2008ca}
Chae, J., Ahn, K., Lim, E.-K., Choe, G.~S., and Sakurai, T.: 2008,
\newblock {\em ApJ} {\bf 689(1)}, L73

\bibitem[\protect\astroncite{Chae et~al.}{2001}]{Chae:2001jl}
Chae, J., Martin, S.~F., Yun, H.~S., Kim, J., Lee, S., Goode, P.~R., Spirock,
  T., and Wang, H.: 2001,
\newblock {\em Astrophysical Journal} {\bf 548(1)}, 497

\bibitem[\protect\astroncite{Chae and Sakurai}{2008}]{Chae:2008jq}
Chae, J. and Sakurai, T.: 2008,
\newblock {\em ApJ} {\bf 689}, 593

\bibitem[\protect\astroncite{Chandrasekhar}{1961}]{Chandrasekhar:1961a}
Chandrasekhar, S.: 1961,
\newblock {\em {Hydrodynamic and Hydromagnetic Stability}},
\newblock Dover

\bibitem[\protect\astroncite{De~Pontieu et~al.}{2014}]{DePontieu:2014aa}
De~Pontieu, B., Title, A.~M., Lemen, J.~R., Kushner, G.~D., Akin, D.~J.,
  Allard, B., Berger, T., Boerner, P., Cheung, M., Chou, C., Drake, J., Duncan,
  D.~W., Freeland, S., Heyman, G.~F., Hoffman, C., Hurlburt, N.~E., Lindgren,
  R.~W., Mathur, D., Rehse, R., Sabolish, D., Seguin, R., Schrijver, C.~J.,
  Tarbell, T.~D., W{\"u}lser, J.~P., Wolfson, C.~J., Yanari, C., Mudge, J.,
  Nguyen-Phuc, N., Timmons, R., van Bezooijen, R., Weingrod, I., Brookner, R.,
  Butcher, G., Dougherty, B., Eder, J., Knagenhjelm, V., Larsen, S., Mansir,
  D., Phan, L., Boyle, P., Cheimets, P.~N., DeLuca, E.~E., Golub, L., Gates,
  R., Hertz, E., McKillop, S., Park, S., Perry, T., Podgorski, W.~A., Reeves,
  K., Saar, S., Testa, P., Tian, H., Weber, M., Dunn, C., Eccles, S., Jaeggli,
  S.~A., Kankelborg, C.~C., Mashburn, K., Pust, N., Springer, L., Carvalho, R.,
  Kleint, L., Marmie, J., Mazmanian, E., Pereira, T. M.~D., Sawyer, S., Strong,
  J., Worden, S.~P., Carlsson, M., Hansteen, V.~H., Leenaarts, J., Wiesmann,
  M., Aloise, J., Chu, K.~C., Bush, R.~I., Scherrer, P.~H., Brekke, P.,
  Mart{\'\i}nez-Sykora, J., Lites, B.~W., McIntosh, S.~W., Uitenbroek, H.,
  Okamoto, T.~J., Gummin, M.~A., Auker, G., Jerram, P., Pool, P., and Waltham,
  N.: 2014,
\newblock {\em Sol Phys} {\bf 289(7)}, 2733

\bibitem[\protect\astroncite{de~Toma et~al.}{2008}]{deToma:2008ei}
de~Toma, G., Casini, R., Burkepile, J.~T., and Low, B.~C.: 2008,
\newblock {\em ApJ} {\bf 687}, L123

\bibitem[\protect\astroncite{Dove et~al.}{2011}]{Dove:2011gh}
Dove, J.~B., Gibson, S.~E., Rachmeler, L.~A., Tomczyk, S., and Judge, P.: 2011,
\newblock {\em ApJ} {\bf 731(1)}, L1

\bibitem[\protect\astroncite{Dud{\'\i}k et~al.}{2012}]{Dudik:2012jy}
Dud{\'\i}k, J., Aulanier, G., Schmieder, B., Zapi{\'o}r, M., and Heinzel, P.:
  2012,
\newblock {\em ApJ} {\bf 761(1)}, 9

\bibitem[\protect\astroncite{Engvold}{1981}]{Engvold:1981aa}
Engvold, O.: 1981,
\newblock {\em Sol Phys} {\bf 70(2)}, 315

\bibitem[\protect\astroncite{Fisher and Welsch}{2008}]{Fisher:2008uf}
Fisher, G.~H. and Welsch, B.~T.: 2008,
\newblock {\em ASP Conference Series} {\bf 383}, 373

\bibitem[\protect\astroncite{Foullon et~al.}{2011}]{Foullon:2011cw}
Foullon, C., Verwichte, E., Nakariakov, V.~M., Nykyri, K., and Farrugia, C.~J.:
  2011,
\newblock {\em The Astrophysical Journal Letters} {\bf 729(1)}, L8

\bibitem[\protect\astroncite{Freed et~al.}{2016}]{Freed:2016gw}
Freed, M.~S., McKenzie, D.~E., Longcope, D.~W., and Wilburn, M.: 2016,
\newblock {\em ApJ} {\bf 818(1)}, 1

\bibitem[\protect\astroncite{Fuller and Gibson}{2009}]{Fuller:2009jm}
Fuller, J. and Gibson, S.~E.: 2009,
\newblock {\em ApJ} {\bf 700}, 1205

\bibitem[\protect\astroncite{Gun{\'a}r et~al.}{2014}]{Gunar:2014he}
Gun{\'a}r, S., Schwartz, P., Dud{\'\i}k, J., Schmieder, B., Heinzel, P., and
  Jur{\v c}{\'a}k, J.: 2014,
\newblock {\em A{\&}A} {\bf 567}, A123

\bibitem[\protect\astroncite{Guzdar et~al.}{1982}]{Guzdar:1982gz}
Guzdar, P.~N., Satyanarayana, P., Huba, J.~D., and Ossakow, S.~L.: 1982,
\newblock {\em Geophys. Res. Lett.} {\bf 9}, 547

\bibitem[\protect\astroncite{Haerendel and Berger}{2011}]{Haerendel:2011de}
Haerendel, G. and Berger, T.: 2011,
\newblock {\em ApJ} {\bf 731(2)}, 82

\bibitem[\protect\astroncite{Hillier}{2017}]{Hillier:2017yy}
Hillier, A.: 2017,
\newblock {\em Rev. Mod. Plasma Phys.},
\newblock accepted

\bibitem[\protect\astroncite{Hillier et~al.}{2012a}]{Hillier:2012fd}
Hillier, A., Berger, T., Isobe, H., and Shibata, K.: 2012a,
\newblock {\em ApJ} {\bf 746(2)}, 120

\bibitem[\protect\astroncite{Hillier et~al.}{2011}]{Hillier:2011hna}
Hillier, A., Isobe, H., Shibata, K., and Berger, T.: 2011,
\newblock {\em ApJ} {\bf 736(1)}, L1

\bibitem[\protect\astroncite{Hillier et~al.}{2012b}]{Hillier:2012bo}
Hillier, A., Isobe, H., Shibata, K., and Berger, T.: 2012b,
\newblock {\em ApJ} {\bf 756(2)}, 110

\bibitem[\protect\astroncite{Hillier et~al.}{2017}]{Hillier:2017go}
Hillier, A., Matsumoto, T., and Ichimoto, K.: 2017,
\newblock {\em A{\&}A} {\bf 597}, A111

\bibitem[\protect\astroncite{Keppens et~al.}{2015}]{Keppens:2015iz}
Keppens, R., Xia, C., and Porth, O.: 2015,
\newblock {\em ApJ} {\bf 806(1)}, 1

\bibitem[\protect\astroncite{Khomenko et~al.}{2014}]{Khomenko:2014br}
Khomenko, E., D{\'\i}az, A., de~Vicente, A., Collados, M., and Luna, M.: 2014,
\newblock {\em A{\&}A} {\bf 565}, A45

\bibitem[\protect\astroncite{Kosugi et~al.}{2007}]{Kosugi:2007ij}
Kosugi, T., Matsuzaki, K., Sakao, T., Shimizu, T., Sone, Y., Tachikawa, S.,
  Hashimoto, T., Minesugi, K., Ohnishi, A., Yamada, T., Tsuneta, S., Hara, H.,
  Ichimoto, K., Suematsu, Y., Shimojo, M., Watanabe, T., Shimada, S., Davis,
  J.~M., Hill, L.~D., Owens, J.~K., Title, A.~M., Culhane, J.~L., Harra, L.~K.,
  Doschek, G.~A., and Golub, L.: 2007,
\newblock {\em Sol Phys} {\bf 243}, 3

\bibitem[\protect\astroncite{Labrosse et~al.}{2010}]{Labrosse:2010bt}
Labrosse, N., Heinzel, P., Vial, J.~C., Kucera, T., Parenti, S., Gun{\'a}r, S.,
  Schmieder, B., and Kilper, G.: 2010,
\newblock {\em Space Sci Rev} {\bf 151(4)}, 243

\bibitem[\protect\astroncite{Leonardis et~al.}{2012}]{Leonardis:2012er}
Leonardis, E., Chapman, S.~C., and Foullon, C.: 2012,
\newblock {\em ApJ} {\bf 745(2)}, 185

\bibitem[\protect\astroncite{Levens et~al.}{2016}]{Levens:2016kz}
Levens, P. J., Schmieder, B., L{\'o}pez Ariste, A., Labrosse, N., Dalmasse, K., and Gelly, B.: 2016,
\newblock{\em ApJ}{\bf 826}, 164

\bibitem[\protect\astroncite{Li et~al.}{2012}]{Li:2012hu}
Li, X., Morgan, H., Leonard, D., and Jeska, L.: 2012,
\newblock {\em ApJ} {\bf 752(2)}, L22

\bibitem[\protect\astroncite{Low}{2001}]{Low:2001vt}
Low, B. C.: 2001,
\newblock {\em Journ. Geophys. Res.}{\bf 106}, A(11), 25141

\bibitem[\protect\astroncite{Liu et~al.}{2012}]{Liu:2012gqa}
Liu, W., Berger, T.~E., and Low, B.~C.: 2012,
\newblock {\em ApJ} {\bf 745(2)}, L21

\bibitem[\protect\astroncite{Low et~al.}{2012a}]{Low:2012tu}
Low, B.~C., Berger, T., and Casini, R.: 2012a,
\newblock {\em ApJ}

\bibitem[\protect\astroncite{Low et~al.}{2012b}]{Low:2012bj}
Low, B.~C., Liu, W., Berger, T., and Casini, R.: 2012b,
\newblock {\em ApJ} {\bf 757(1)}, 21

\bibitem[\protect\astroncite{Mackay et~al.}{2010}]{Mackay:2010fp}
Mackay, D.~H., Karpen, J.~T., Ballester, J.~L., Schmieder, B., and Aulanier,
  G.: 2010,
\newblock {\em Space Sci Rev} {\bf 151(4)}, 333

\bibitem[\protect\astroncite{Marsch et~al.}{2008}]{Marsch:2008aa}
Marsch, E., Tian, H., Sun, J., Curdt, W., and Wiegelmann, T.: 2008,
\newblock {\em ApJ} {\bf 685(2)}, 1262

\bibitem[\protect\astroncite{McIntosh et~al.}{2012}]{McIntosh:2012dx}
McIntosh, S.~W., Tian, H., Sechler, M., and De~Pontieu, B.: 2012,
\newblock {\em ApJ} {\bf 749(1)}, 60

\bibitem[\protect\astroncite{M{\"o}stl et~al.}{2013}]{Mostl:2013ir}
M{\"o}stl, U.~V., Temmer, M., and Veronig, A.~M.: 2013,
\newblock {\em ApJ Letters} {\bf 766(1)}, L12

\bibitem[\protect\astroncite{Ofman and Thompson}{2011}]{Ofman:2011gs}
Ofman, L. and Thompson, B.~J.: 2011,
\newblock {\em ApJ} {\bf 734(1)}, L11

\bibitem[\protect\astroncite{Okamoto et~al.}{2016}]{Okamoto:2016tw}
Okamoto, T.~J., Liu, W., and Tsuneta, S.: 2016,
\newblock {\bf 831(2)}, 126

\bibitem[\protect\astroncite{Okamoto et~al.}{2010}]{Okamoto:2010gi}
Okamoto, T.~J., Tsuneta, S., and Berger, T.~E.: 2010,
\newblock {\em ApJ} {\bf 719}, 583

\bibitem[\protect\astroncite{Okamoto et~al.}{2009}]{Okamoto:2009cg}
Okamoto, T.~J., Tsuneta, S., Lites, B.~W., Kubo, M., Yokoyama, T., Berger,
  T.~E., Ichimoto, K., Katsukawa, Y., Nagata, S., Shibata, K., Shimizu, T.,
  Shine, R.~A., Suematsu, Y., Tarbell, T.~D., and Title, A.~M.: 2009,
\newblock {\em ApJ} {\bf 697}, 913

\bibitem[\protect\astroncite{Olson et~al.}{2011}]{Olson:2011ht}
Olson, B.~J., Larsson, J., Lele, S.~K., and Cook, A.~W.: 2011,
\newblock {\em Phys. Fluids} {\bf 23(11)}, 114107

\bibitem[\protect\astroncite{Orozco~Suarez et~al.}{2014}]{OrozcoSuarez:2014he}
Orozco~Suarez, D., Asensio~Ramos, A., and Trujillo~Bueno, J.: 2014,
\newblock {\em A{\&}A} {\bf 566}, A46

\bibitem[\protect\astroncite{Panesar et~al.}{2013}]{Panesar:2013gu}
Panesar, N.~K., Innes, D.~E., Tiwari, S.~K., and Low, B.~C.: 2013,
\newblock {\em A{\&}A} {\bf 549}, A105

\bibitem[\protect\astroncite{Parenti et~al.}{2012}]{Parenti:2012cc}
Parenti, S., Schmieder, B., Heinzel, P., and Golub, L.: 2012,
\newblock {\em ApJ} {\bf 754(1)}, 66

\bibitem[\protect\astroncite{Ruderman et~al.}{2014}]{Ruderman:2014cq}
Ruderman, M.~S., Terradas, J., and Ballester, J.~L.: 2014,
\newblock {\em ApJ} {\bf 785(2)}, 110

\bibitem[\protect\astroncite{Ryutova et~al.}{2010}]{Ryutova:2010fd}
Ryutova, M., Berger, T., Frank, Z., Tarbell, T., and Title, A.: 2010,
\newblock {\em Sol Phys} p. 170

\bibitem[\protect\astroncite{Schmieder et~al.}{2014}]{Schmieder:2014kt}
Schmieder, B., Tian, H., Kucera, T., López Ariste, A., Mein, N., Mein, P., Dalmasse, K., and Golub, L.: 2014,
\newblock{\em Astron. Astrophys.} {\bf 569}, A85

\bibitem[\protect\astroncite{Schmit and Gibson}{2011}]{Schmit:2011ix}
Schmit, D.~J. and Gibson, S.~E.: 2011,
\newblock {\em ApJ} {\bf 733(1)}, 1

\bibitem[\protect\astroncite{Shen et~al.}{2015}]{Shen:2015gx}
Shen, Yuandeng and Liu, Yu and Liu, Ying D and Chen, P F and Su, Jiangtao and Xu, Zhi and Liu, Zhong: 2015,
\newblock {\em ApJ} {\bf 814}, L17

\bibitem[\protect\astroncite{Stellmacher and Wiehr}{1973}]{Stellmacher:1973vg}
Stellmacher, G. and Wiehr, E.: 1973,
\newblock {\em A{\&}A} {\bf 24}, 321

\bibitem[\protect\astroncite{Stone and Gardiner}{2007a}]{Stone:2007jx}
Stone, J.~M. and Gardiner, T.: 2007a,
\newblock {\em Phys. Fluids} {\bf 19}, 4104

\bibitem[\protect\astroncite{Stone and Gardiner}{2007b}]{Stone:2007cw}
Stone, J.~M. and Gardiner, T.: 2007b,
\newblock {\em ApJ} {\bf 671}, 1726

\bibitem[\protect\astroncite{Suematsu et~al.}{2008}]{Suematsu:2008gq}
Suematsu, Y., Tsuneta, S., Ichimoto, K., Shimizu, T., Otsubo, M., Katsukawa,
  Y., Nakagiri, M., Noguchi, M., Tamura, T., Kato, Y., Hara, H., Kubo, M.,
  Mikami, I., Saito, H., Matsushita, T., Kawaguchi, N., Nakaoji, T., Nagae, K.,
  Shimada, S., Takeyama, N., and Yamamuro, T.: 2008,
\newblock {\em Sol Phys} {\bf 249}, 197

\bibitem[\protect\astroncite{Terradas et~al.}{2015}]{Terradas:2015jn}
Terradas, J., Soler, R., Luna, M., Oliver, R., and Ballester, J.~L.: 2015,
\newblock {\em ApJ} {\bf 799(1)}, 94

\bibitem[\protect\astroncite{Tsuneta et~al.}{2008}]{Tsuneta:2008be}
Tsuneta, S., Ichimoto, K., Katsukawa, Y., Nagata, S., Otsubo, M., Shimizu, T.,
  Suematsu, Y., Nakagiri, M., Noguchi, M., Tarbell, T., Title, A., Shine, R.,
  Rosenberg, W., Hoffmann, C., Jurcevich, B., Kushner, G., Levay, M., Lites,
  B., Elmore, D., Matsushita, T., Kawaguchi, N., Saito, H., Mikami, I., Hill,
  L.~D., and Owens, J.~K.: 2008,
\newblock {\em Sol Phys} {\bf 249(2)}, 167

\bibitem[\protect\astroncite{Vial and Engvold}{2015}]{Vial:2015cn}
Vial, J. C. and Engvold, O., Ed.: 2015,
\newblock {\em Solar Prominences}, Springer, ISSN 0067-0057

\bibitem[\protect\astroncite{Xia and Keppens}{2016}]{Xia:2016vs}
Xia, C. and Keppens, R.: 2016,
\newblock {\em ApJ} {\bf 823(1)}, 22

\bibitem[\protect\astroncite{Ye et~al.}{2011}]{Ye:2011gl}
Ye, W.~H., Wang, L.~F., Xue, C., Fan, Z.~F., and He, X.~T.: 2011,
\newblock {\em Phys. Plasmas} {\bf 18(2)}, 022704

\bibitem[\protect\astroncite{Zhang et~al.}{2005}]{Zhang:2005it}
Zhang, W., Wu, Z., and Li, D.: 2005,
\newblock {\em Phys. Plasmas} {\bf 12(4)}, 042106

\end{thebibliography}
\end{document}